\documentclass{aa}  

\usepackage{graphicx}
\usepackage{txfonts}
\usepackage{lipsum}
\usepackage{lscape}             
\usepackage{placeins}           
                                
\usepackage{hyperref}
\usepackage{multirow}
\usepackage[dvipsnames]{xcolor}
\usepackage{rotating}
\usepackage{booktabs}

\begin{document}

   \title{Survey of Surveys Data Release 3. Homogenised stellar parameters across nine spectroscopic surveys}


   \author{A. Avdeeva\inst{\ref{oaa}}
        \and E. Pancino\inst{\ref{oaa}}
        \and A.~Turchi\inst{\ref{oaa}} \and P.M.~Marrese\inst{\ref{oar},\ref{ssdc}}
        \and    S.~Marinoni\inst{\ref{oar},\ref{ssdc}}
        \and G.~Fanari\inst{\ref{ssdc}}
        \and D.~Alvarez Garay\inst{\ref{oaa}} \and 
        M. Echeveste\inst{\ref{oaa}}\and 
        E. Leitinger\inst{\ref{oaa} \and \ref{unibo}} \and
        S. Nedhath\inst{\ref{oaa}\and \ref{unifi}}\and 
        S. Rani\inst{\ref{oaa}}\and 
        E. Reggiani\inst{\ref{oaa}, \ref{unifi}}\and 
        N.~Sanna\inst{\ref{oaa}}\and 
        S. Saracino\inst{\ref{oaa}}\and 
        L. Steinbauer\inst{\ref{oaa}, \ref{unifi}}
        }

   \institute{INAF -- Osservatorio Astrofisico di Arcetri, Largo E. Fermi 5, 50125 Firenze, Italy\label{oaa}
\and Space Science Data Center, Via del Politecnico SNC, I-00133 Roma, Italy\label{ssdc}
\and INAF -- Osservatorio Astronomico di Roma, Via Frascati 33, 00040, Monte Porzio Catone, Roma, Italy\label{oar}
\and Dipartimento di Fisica e Astronomia, Università di Firenze,  Via G. Sansone 1, 50019 Sesto Fiorentino, FI, Italy\label{unifi} \and Dipartimento di Fisica e Astronomia, Universit{\`a} degli Studi di Bologna, Via Gobetti 93/2, 40129 Bologna, Italy\label{unibo}}
   \date{Received XX, 20XX}

 
\abstract
   {Large spectroscopic surveys have measured atmospheric parameters for
millions of Milky Way stars. However, these surveys rely on different
instruments, wavelength ranges, analysis methods, and assumptions. Their
published stellar parameters are therefore affected by survey-dependent
systematics, which makes their joint use difficult, especially for studies
that combine data from several surveys without explicitly accounting for
these offsets.}
   {We aim to place effective temperature, surface gravity, and metallicity
estimates from major spectroscopic surveys onto a well-calibrated common
scale, externally validated against high-quality reference data, and to
provide a homogenised catalogue suitable for large-scale Galactic studies.}
   {We construct the calibration hierarchically. First, we define an internal
reference scale from the best-populated and best-overlapped surveys:
APOGEE, GALAH, and LAMOST LRS. We then model survey-to-reference
corrections and their uncertainties with machine-learning regression. The
resulting scale is extended to LAMOST MRS, \textit{Gaia} RVS, RAVE, DESI,
\textit{Gaia}-ESO, and SEGUE, and is finally anchored to high-quality measurements
from the PASTEL database.}
   {The calibration reduces survey-to-survey offsets and improves the
   consistency of the stellar parameters in independent overlap samples.
   Independent calibration of GALAH and LAMOST LRS temperatures onto the
   APOGEE scale reduces their median offset from 44.5\,K to 6.2\,K. The final catalogue contains approximately 10.13 million stars, including
6.10 million stars in the recommended high-quality subset and 3.69 million
stars in the metallicity-optimised subset. For the recommended high-quality subset, validation with \textit{Gaia} benchmark stars gives
    median residuals of 30\,K in $T_{\rm eff}$, 0.03\,dex in $\log g$,
    and -0.02\,dex in [Fe/H]. For metallicity-focused applications, the
    [Fe/H]-optimised subset gives median residuals of 0.003\,dex against
    \textit{Gaia} benchmark stars, 0.05\,dex for globular clusters, and -0.02\,dex
    for open clusters.
   As an example science application, we use the homogenised catalogue
to confirm the presence of metal-poor stars on thin-disc-like orbits,
including stars with $\mathrm{[Fe/H]} < -2$, showing that the catalogue
can be used to study rare Galactic populations on a consistent
multi-survey parameter scale.}
   {}


   \maketitle
\nolinenumbers


\section{Introduction}
\label{sec:intro}

Large spectroscopic surveys have transformed stellar and Galactic astrophysics over the past two decades. Projects such as APOGEE\footnote{\url{https://www.sdss.org/surveys/apogee/}},
GALAH\footnote{\url{https://www.galah-survey.org/}},
LAMOST\footnote{\url{https://www.lamost.org/}},
RAVE\footnote{\url{https://www.rave-survey.org/}},
\textit{Gaia} RVS\footnote{\url{https://www.cosmos.esa.int/web/gaia/dr3}},
and DESI\footnote{\url{https://www.desi.lbl.gov/}}
have measured atmospheric parameters for millions of stars across different
regions of the Milky Way
\citep[e.g.][]{Majewski2017,deSilva2015,Zhao2012,Steinmetz2006,RecioBlanco2023,DESICollaboration2016a}.

These surveys probe complementary magnitude ranges, Galactic components, and evolutionary stages, providing an unprecedented view of the structure and chemical evolution of our Galaxy. However, fully benefiting from the scientific potential of these surveys is difficult when their parameter scales are used together without explicit calibration.
Differences in wavelength coverage, spectral resolution, target selection, analysis methods, and underlying assumptions introduce survey-dependent zero-points and parameter-dependent trends in the reported effective temperatures ($T_{\rm eff}$), surface gravities ($\log g$), and metallicities ([Fe/H]). These effects can propagate into the inferred structure and chemo-dynamical properties of the Milky Way, especially when studies rely on stars drawn from several surveys or focus on sparsely populated regions of parameter space. A homogeneous parameter scale is therefore needed not only for catalogue-level consistency, but also for robust comparisons of Galactic populations across different observational regimes.

Several works have addressed cross-survey calibration, increasingly using data-driven ``label transfer'' trained on overlap samples to place different surveys on one common scale, including The Cannon \citep{Ness2015, Casey2016, Ho2017}, The Payne \citep{Ting2019}, and StarNet \citep{Fabbro2018,Bialek2020}, with further multi-survey examples discussed in the same context \citep[e.g.][]{Xiang2019,Wheeler2020,Nandakumar2022}. In this wider literature, \citet{Thomas2024} proposed a neural-network method that translates pipeline outputs from one survey base to another (e.g., GALAH to APOGEE), treating the differences as a multi-parameter, non-linear transformation instead of a single global offset, and stressing that the method is most reliable where the overlap sample provides good coverage. \citet{Guiglion2024} provided a complementary example in which stellar parameters and abundances for \textit{Gaia} RVS are inferred using \textit{Gaia} \citep{Prusti2016} spectra together with additional \textit{Gaia} information, and the results are placed on the APOGEE reference scale, showing how anchoring to an external reference can be built directly into the inference workflow. Finally, large collaborations highlight that it is also important to control internal differences within a survey before comparing to others, as shown by the \textit{Gaia}-ESO strategy for homogenisation and calibration across analysis pipelines \citep{Pancino2017}.

This work continues the Survey of Surveys (SoS) series by constructing
a unified and publicly available catalogue of homogenised stellar
parameters from major spectroscopic surveys. In SoS-I \citep{Tsantaki2022}, we constructed the largest homogenised radial velocity (RV) catalogue available at the time, placing \textit{Gaia} and major ground-based spectroscopic surveys onto a common reference scale. The procedure included cross-matching, uncertainty renormalisation, parameter-dependent trend corrections and validation against external standards, delivering RVs for nearly 11 million stars.

SoS-II \citep{Turchi2025} represented a major extension of SoS-I in two aspects. First, the homogenisation framework was extended from RVs to spectroscopically derived stellar atmospheric parameters ($T_{\rm eff}$, $\log g$, and [Fe/H]). Five spectroscopic surveys were internally and externally calibrated onto a common scale anchored to high-resolution reference data. Second, this calibrated spectroscopic scale was used as a training reference to infer atmospheric parameters from large photometric surveys Sloan Digital Sky Survey \citep[SDSS;][]{Abazajian03}, the SkyMapper Southern Sky Survey \citep[SM;][]{Keller07} with machine-learning methods. This second step produced homogenised stellar parameters for about 23 million stars and demonstrated substantial improvements over previous similar works, particularly in the metal-poor regime ([Fe/H]<-1).

In this work, we construct a unified compilation of stellar atmospheric parameters by placing nine major spectroscopic surveys onto a common internal reference scale and subsequently calibrating this scale to an external standard (PASTEL). Our approach combines empirical error re-normalisation, data-driven regression models, and uncertainty calibration techniques to ensure both accuracy and statistical consistency. The resulting catalogue enables cross-survey analyses over a wide range of magnitudes, sky coverage, and stellar populations.

This paper is organized as follows. In Section~\ref{sec:data}, we describe the input
surveys and the applied quality filters. Section~\ref{sec:method} presents the calibration
methodology, including uncertainty re-scaling, machine-learning-based
corrections, and internal validation. In Section~\ref{sec:validation}, we perform the external
calibration and validation using PASTEL, \textit{Gaia} benchmark stars, APOKASC, and
stellar clusters. Section~\ref{sec:catalogue} describes the final catalogue and defines the
high-quality and metallicity-optimised samples. Section~\ref{sec:sci_case} presents an example
science application to metal-poor stars on thin-disc-like orbits. We
summarize our results in Section~\ref{sec:conclusion}.

\section{Data}
\label{sec:data}

This work combines $T_{\mathrm{eff}}$, $\log g$, and $[\mathrm{Fe}/\mathrm{H}]$ from the major spectroscopic surveys currently used in large-sample studies of Milky Way stars.  In this section, we describe the input datasets and the quality filtering applied before homogenisation.

\subsection{Surveys description}

\begin{table*}
\caption{Summary of spectroscopic surveys used in this work.}
\centering
\begin{tabular}{lcccc}
\hline\hline
Survey & No. of entries & Wavelength Coverage & Resolution ($R$) \\
\hline
APOGEE DR17 & $\sim$727\,000 & 1.51--1.69\,$\mu$m & $\sim$22\,500  \\
\multirow{2}{*}{GALAH DR4} & \multirow{2}{*}{$\sim$917\,000} & 4713--4903\,\AA , 5648--5873\,\AA , & \multirow{2}{*}{$\sim$28\,000} \\
 &  & 6478--6737\,\AA , 7585--7887\,\AA  &  &  \\
LAMOST DR11 (LRS) & $\sim$7.9 million (AFGK stars) & 3700--9000\,\AA  & $\sim$1800  \\
LAMOST DR11 (MRS) & $\sim$2.6 million & 4950--5350\,\AA , 6300--6800\,\AA & $\sim$7500  \\
\textit{Gaia} RVS (DR3) & $\sim$5.6 million (1.8 million with [Fe/H]) & 8470--8740\,\AA  & $\sim$11\,500\\
RAVE DR6 & $\sim$450\,000 & 8410--8795\,\AA  & $\sim$7500 \\
DESI DR1 (stellar VAC) & $\sim$6.4 million & 3600--9800\,\AA & $\sim$2000--5000  \\
\textit{Gaia}-ESO DR5 & $\sim$115\,000 & $\sim$4800--8750\,\AA \tablefootmark{a} & $\sim$20\,000--47\,000 \\
SEGUE (1, 2 and legacy) & $\sim$440\,000 & $\sim$3800--9200\,\AA & $\sim$1800  \\
\hline
\end{tabular}
\tablefoot{
\tablefoottext{a}{Coverage varies depending on GIRAFFE or UVES setup.}}
\label{tab:catalogue_summary}
\end{table*}

The survey set used in this work consists of the large spectroscopic datasets currently available for stellar-parameter studies: APOGEE, GALAH, LAMOST, \textit{Gaia} RVS, RAVE, DESI, \textit{Gaia}-ESO and SEGUE. Their combination is motivated primarily by practical relevance: they provide the broadest basis for a public homogenised catalogue. The number of entries contributed by each survey as well as details on wavelength coverage and resolution are listed in Table~\ref{tab:catalogue_summary}.

\begin{figure*}[ht!]
    \centering
    \includegraphics[width=0.85\textwidth]{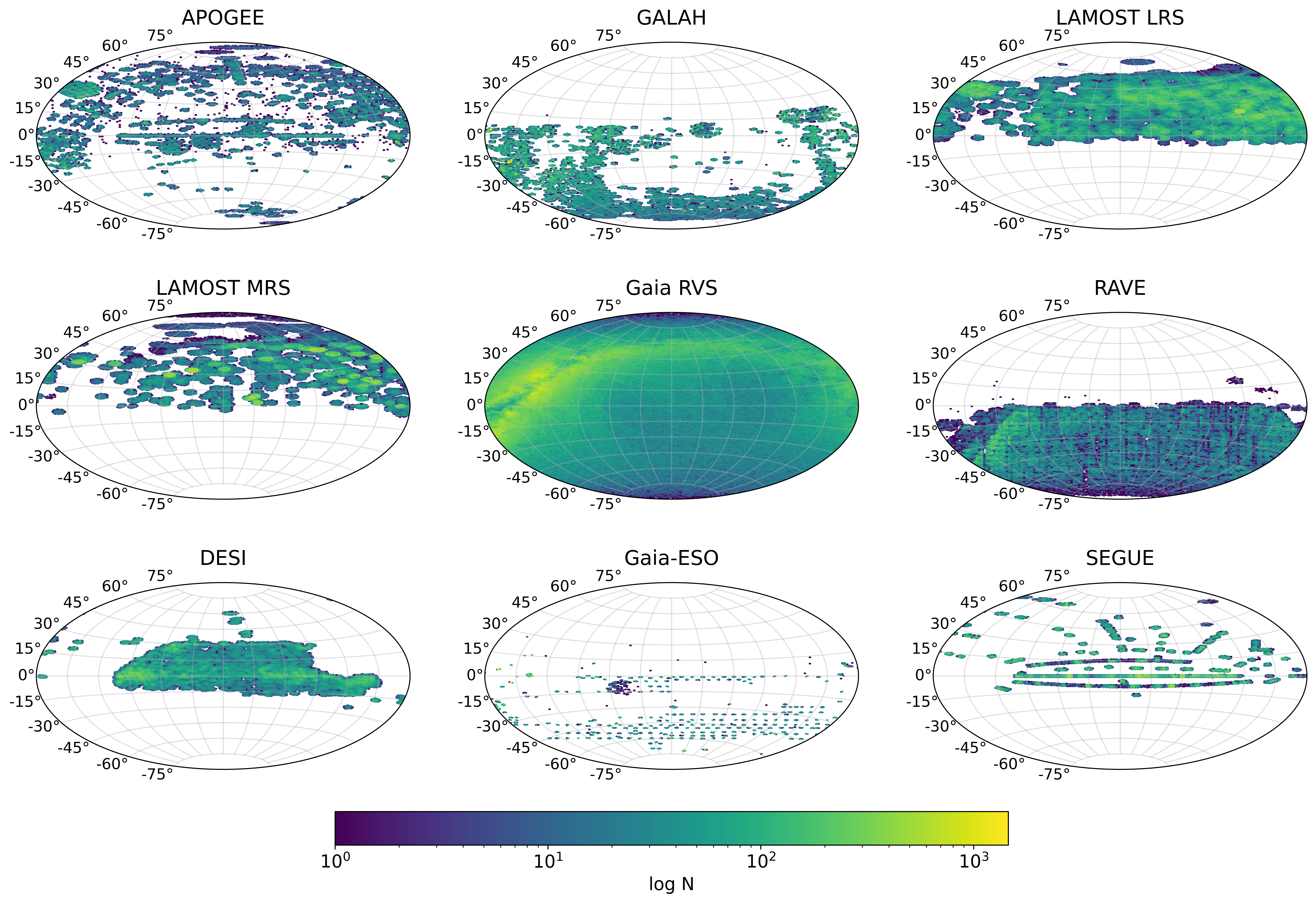}
    \caption{Aitoff projection of the RA-Dec distributions for all the surveys used in this work.}
    \label{fig:aitoff}
\end{figure*}

The seventeenth APOGEE data release (DR17; \citealt{Abdurrouf2022}) provides high-resolution near-infrared spectroscopy obtained with the APOGEE spectrographs on the Sloan 2.5 m telescopes in both hemispheres. Observing in the $H$ band, APOGEE primarily targets red giant stars over approximately $7 \lesssim H \lesssim 13.8$ mag, enabling studies of distant and dust-obscured regions of the Galactic disk and bulge. $T_{\mathrm{eff}}$, $\log g$, and [Fe/H] are derived with the APOGEE Stellar Parameter and Chemical Abundances Pipeline \citep[ASPCAP]{GarciaPerez2016}, based on global spectral fitting.

The DR4 of GALAH survey \citep{Buder2025} provides high-resolution optical spectroscopy obtained with the HERMES spectrograph \citep{deSilva2012} on the 3.9 m Anglo-Australian Telescope. GALAH primarily targets bright stars in the southern Galactic disk, typically in the range $12 \lesssim V \lesssim 14$ mag, and is optimised for detailed chemical characterisation of nearby stellar populations. In DR4, stellar labels are derived iteratively by combining spectroscopic fitting with astrometric and photometric constraints, and the release provides atmospheric parameters and detailed elemental abundances.

LAMOST DR11 provides wide-area spectroscopic coverage of the northern sky in both low-resolution (LRS) and medium-resolution (MRS) modes. The LRS data provide broad wavelength coverage and sample both disk and halo populations over a wide magnitude range, typically $r=9$--17.8 mag. The MRS data provide higher-resolution spectra mainly for brighter targets, including time-domain and non-time-domain programmes. Stellar atmospheric parameters are provided by the LAMOST Stellar Parameter Pipeline \citep[LASP]{Wu2014}. Although DR11 also includes convolutional neural network-based parameters for MRS combined spectra, these are not accompanied by published per-object uncertainties in the current release. We therefore adopt the LASP determinations and treat the LRS and MRS samples as separate datasets.

RAVE DR6 \citep{Steinmetz2020a, Steinmetz2020b} is the final data release of the Radial Velocity Experiment, based on medium-resolution spectroscopy of the near-infrared calcium triplet region. The survey primarily targeted southern-hemisphere field stars in the range $9 \lesssim I \lesssim 12$ mag, bridging nearby and intermediate-distance disk populations. Stellar parameters are derived with the updated MADERA pipeline, which provides $T_{\mathrm{eff}}$, $\log g$, and global metallicity [M/H], while individual iron abundance [Fe/H] estimates are available for a subset of the sample through an auxiliary table.

The ESA \textit{Gaia} mission’s Radial Velocity Spectrometer \citep[RVS]{Katz2004, Cropper2018} provides all-sky medium-resolution spectroscopy for stars down to approximately $G \lesssim 16$ mag, although the effective limit for reliable atmospheric-parameter determination depends on signal-to-noise and crowding. In \textit{Gaia} DR3 \citep{GaiaCollaboration2023}, atmospheric parameters are derived by the General Stellar Parametriser from spectroscopy \citep[GSP-Spec]{RecioBlanco2023}, which provides $T_{\mathrm{eff}}$, $\log g$, overall metallicities [M/H], and [$\alpha$/Fe] ratios for several million stars.

The DESI stellar Value-Added Catalogue \citep{Koposov2026} is based on wide-area optical spectroscopy from DESI DR1 \citep{DESICollaboration2025}, covering $\sim1.5$--$2\times10^4$ deg$^2$ of the northern sky and targeting stars over $G\approx12$--21 mag. DESI therefore provides stellar parameters for stars significantly fainter than most of the other surveys considered here, based on low-resolution spectra. Atmospheric parameters are derived using two independent pipelines: \texttt{RVSpecFit} \citep{Koposov2019}, which forward-models spectra using PHOENIX-based templates and a neural-network emulator (RV pipeline), and the SP pipeline, based on \texttt{FERRE} \citep{AllendePrieto2015}, which fits continuum-normalised spectra with synthetic model grids.

The DR5 of \textit{Gaia}-ESO Survey \citep{Gilmore2022, Hourihane2023} provides high-resolution spectroscopy obtained with the FLAMES multi-object spectrograph on the ESO VLT. The survey targeted both field stars and members of open and globular clusters, extending to $V\sim19$ mag for some setups and covering a broad range of evolutionary stages and metallicities. Parameter determination relies on multi-node analyses within dedicated Working Groups, followed by survey-wide homogenisation and calibration.

SEGUE \citep{Yanny2009,Rockosi2022} and related SDSS stellar programmes provide large-area, low-resolution optical spectroscopy of Milky Way stars in a magnitude-limited regime, roughly $14 \lesssim g \lesssim 20$ mag, across many discrete lines of sight. Stellar parameters ($T_{\rm eff}$, $\log g$, and metallicity) are derived with the SEGUE Stellar Parameter Pipeline (SSPP), which combines multiple independent estimators, including spectral matching and line-index methods with auxiliary photometric constraints, into a single adopted solution.

Owing to their different magnitude limits, instrumental configurations, sky coverages, and target-selection functions, these surveys probe complementary Galactic volumes and stellar populations. Their combination therefore provides broader coverage of the Milky Way in both spatial and atmospheric-parameter space than any individual survey alone.

Finally, PASTEL \citep{Soubiran2010,Soubiran2016} is a bibliographic compilation of published $T_{\rm eff}$, $\log g$, and [Fe/H] derived from detailed analyses of high-resolution ($R \geq 25\,000$), high signal-to-noise (S/N $\geq 50$) spectra. In this work, it is used as an external literature reference scale (see also Appendix~\ref{app:pastel_calibration}).


\subsection{Quality filtering}
\label{subsec:quality}

\begin{table*}[h!]
\centering
\caption{Quality cuts applied to the input spectroscopic catalogues before homogenisation.}
\label{tab:quality_filters}
\begin{tabular}{lcc}
\hline
\hline
Survey & Applied cut & N (\%) \\
\hline
APOGEE DR17 & Excluded stars with the \texttt{STAR\_WARN} bit set in \texttt{ASPCAPFLAG} & 624\,640 (86\%) \\
GALAH DR4 & \texttt{flag\_sp} $=0$ and \texttt{snr\_px\_ccd3}$>30$ & 631\,925 (69\%)\\
LAMOST DR11 LRS & \texttt{snrr} $>30$ and \texttt{snrg} $>30$ & 4\,383\,182 (55\%) \\
LAMOST DR11 MRS & \texttt{teff\_lasp}, \texttt{logg\_lasp}, \texttt{feh\_lasp} $\neq -9999$ & 2\,592\,439 (99\%)\\
\textit{Gaia} RVS DR3 & no ``9'' in the first 13 digits of \texttt{flags\_gspspec} and non-null \texttt{fem\_gspspec} & 1\,774\,698 (\bf{32\%})\\
RAVE DR6 & \texttt{algo\_conv\_madera }$=0$ or \texttt{snr\_madera} $\geq 40$ & 363\,814 (81\%)\\
DESI DR1 & \texttt{FEH} $>-3.5$, \texttt{LOGG} $<5.5$, \texttt{TEFF} $<8000$ K, and (\texttt{SUCCES}S $=1$ or \texttt{SNR\_MED} $>50$) & 3\,921\,434 (62\%)\\
\textit{Gaia}-ESO DR5 & \texttt{SFLAGS} not containing neither SRP nor IPA & 79\,932 (70\%)\\
SEGUE 1\&2 & \texttt{FLAG} = nnnnn; excluded GAL, QSO, SKY, WD, PHO, RED, HOT, ROS, QA targets & 259\,282 (59\%)\\
\hline
\end{tabular}

\tablefoot{
\texttt{ASPCAPFLAG} is the APOGEE Stellar Parameter and Chemical Abundances Pipeline quality flag.
\texttt{flag\_sp} is the GALAH spectroscopic quality flag.
\texttt{snr\_px\_ccd3}, \texttt{snrr}, \texttt{snrg}, \texttt{snr\_madera}, and \texttt{SNR\_MED} are signal-to-noise indicators in the corresponding catalogues.
\texttt{flags\_gspspec} is the \textit{Gaia} GSP-Spec processing flag. \texttt{SFLAGS} are the simplified \textit{Gaia}-ESO flags: SRP indicates problematic reduction and IPA incomplete stellar parameters.
For SEGUE, \texttt{FLAG} is the SSPP quality flag, while the excluded
\texttt{TARGET} labels identify objects selected as non-stellar sources,
sky fibres, calibration targets, or other special target classes rather than
standard stellar-population targets: galaxies (GAL), quasars
(QSO), sky fibres (SKY), white-dwarf candidates
(WD), photometric standards (PHO), reddening standards
(RED), hot-standard targets (HOT), ROSAT-selected targets
(ROS), and quality-assurance targets (QA). When an explicit S/N threshold is listed, it is applied directly to the survey data. For LAMOST MRS, no additional S/N cut is imposed because the LASP atmospheric parameters are released only for spectra with S/N>10 in the blue or red band used for parameter determination. For Gaia RVS, GSP-Spec analysed spectra were pre-selected to have S/N>20 before resampling. For Gaia-ESO, no S/N-related criterion is included in our adopted quality selection. For SEGUE, we exclude spectra flagged by the SSPP as very noisy, but we do not impose a specific numerical S/N threshold. N(\%) gives the number of stars retained after the quality cuts and the corresponding fraction of the initial survey sample.
}

\end{table*}

Before homogenisation, we applied survey-specific quality cuts, summarized in
Table~\ref{tab:quality_filters}. These cuts follow the quality flags and
recommendations of the original catalogues whenever available; otherwise, we
adopted conservative empirical filters based on parameter distributions,
overlap samples, and validation diagnostics. The aim was to remove clearly
unreliable stellar-parameter estimates rather than to define a homogeneous
astrophysical selection.

\section{Method}
\label{sec:method}

We construct the calibration hierarchically. In the first stage, we use APOGEE DR17, GALAH DR4 and LAMOST LRS DR11 to define the initial internal reference scale. APOGEE and GALAH provide the highest-resolution large-survey parameter sets in our compilation, while LAMOST LRS supplies a larger overlap sample and repeated observations that are used to empirically reassess the reported uncertainties. Although LAMOST MRS has higher spectral resolution, we prioritise LRS because its substantially larger overlaps with APOGEE and GALAH provide a stronger connection to the initial reference scale. LAMOST LRS also offers larger calibration samples for RAVE and DESI, including better coverage of the metal-poor regime.

The cross-match between surveys was performed through \textit{Gaia} source
identifiers. For APOGEE, RAVE DR6, and \textit{Gaia}-ESO, we used the \textit{Gaia}
cross-matches prepared within the Survey of Surveys framework. These
cross-matches follow the strategy adopted in SoS-I \citep{Tsantaki2022},
based on the \textit{Gaia} external-catalogue cross-match methodology described by
\citet{Marrese2017,Marrese2019}. For APOGEE, the cross-match used in
SoS-I was originally constructed for APOGEE DR16 and was subsequently
updated to APOGEE DR17 for the present work. For GALAH DR4, we adopted the \textit{Gaia} identifiers provided by the GALAH
team. For LAMOST LRS, LAMOST MRS and DESI, we used the \textit{Gaia}
cross-identifications distributed with each survey product.
For SEGUE, we adopted the official \textit{Gaia} DR3--SDSS DR13 cross-match
provided in the \textit{Gaia} Archive through the
\texttt{sdssdr13\_best\_neighbour} table\footnote{\href{https://gea.esac.esa.int/archive/documentation/GDR3/Gaia_archive/chap_datamodel/sec_dm_cross-matches/ssec_dm_sdssdr13_best_neighbour.html}{\textit{Gaia} DR3 - SDSS best neighbour table}},
following the \textit{Gaia} archive cross-match procedure
\citep{Marrese2017,Marrese2019}.
 All
survey-to-survey overlaps used for calibration were then constructed using
the resulting \textit{Gaia} source identifiers. Objects without \textit{Gaia} source identifiers are treated separately during the
final catalogue construction and not used in the calibration process itself. The coordinate and parameter-based merging
procedure used for these sources is described in Appendix~\ref{app:no_gaia}.

\begin{figure}[ht]
    \centering        \includegraphics[width=\linewidth]        {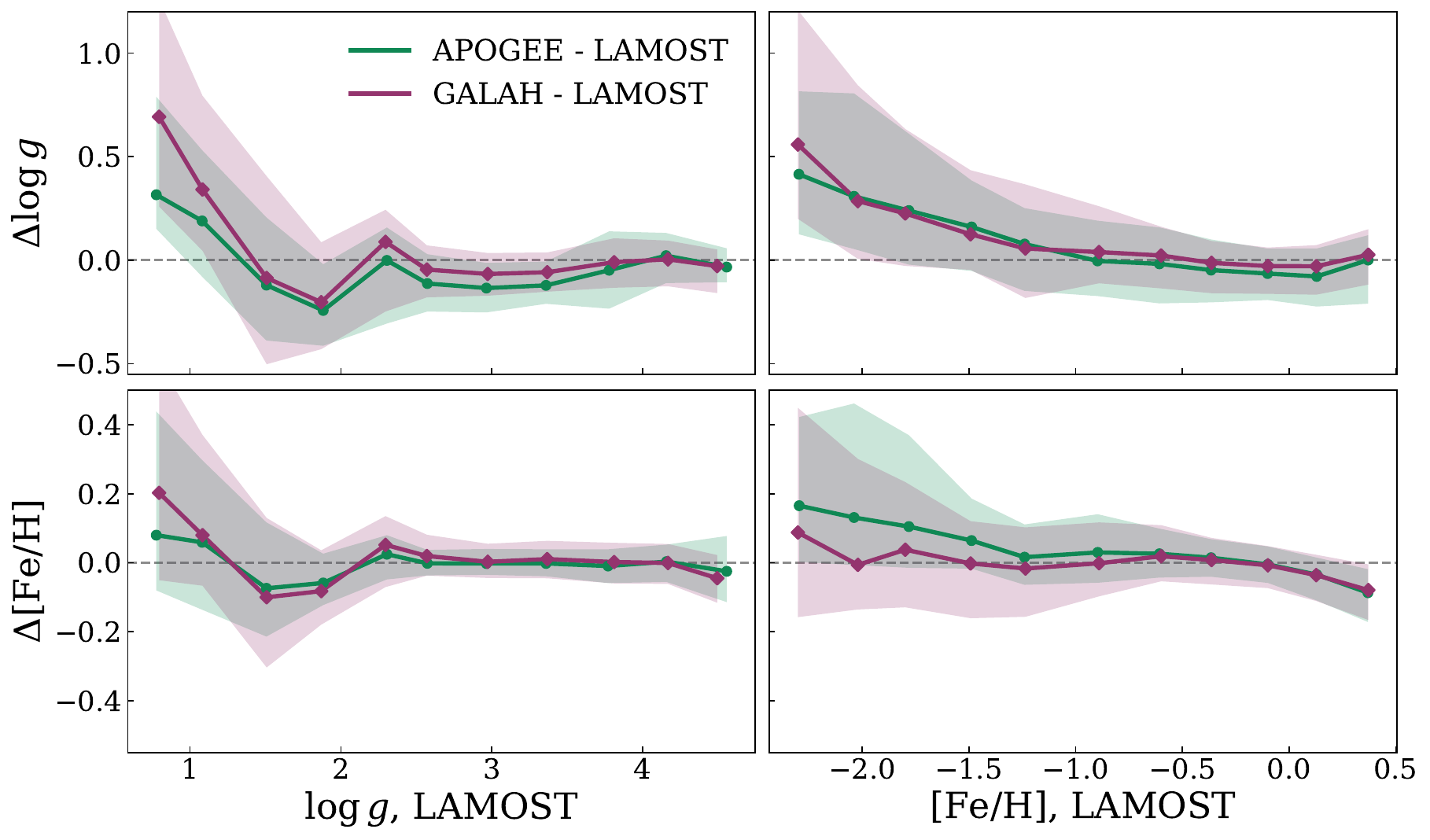}
    \caption{
    Binned differences between APOGEE and LAMOST LRS (green) and between GALAH and LAMOST LRS
    (magenta) for $\log g$ and $[\mathrm{Fe/H}]$. The offsets are shown as functions of
    the LAMOST LRS $\log g$ and $[\mathrm{Fe/H}]$, respectively. 
    The figure illustrates the systematic trends between LAMOST LRS and the two reference surveys before applying the calibration.    }
\label{fig:feh_logg_lamost_trend}
\end{figure}

Pairwise comparisons in the overlap sample between APOGEE, GALAH and LAMOST LRS are used to define how the internal reference scale is constructed. These comparisons reveal systematic offsets and parameter-dependent trends and show that the three atmospheric parameters do not admit a single uniform calibration strategy. In particular, in the LAMOST LRS-APOGEE/GALAH overlap, the [Fe/H] residuals show coherent trends with LAMOST LRS parameters, particularly with $\log g$ and [Fe/H], indicating that the LAMOST LRS metallicity scale can be mapped onto the APOGEE/GALAH reference through a parameter-dependent correction. Similar behaviour is observed in the $\log g$ residuals; see Fig.~\ref{fig:feh_logg_lamost_trend} for both examples. 

For $T_{\rm eff}$, we do not define the internal reference scale in the
same way as for $\log g$ and [Fe/H]. In the APOGEE--GALAH--LAMOST
overlap, the three surveys show good agreement only over a limited
temperature interval, while their scales diverge outside the main FGK
regime. We therefore avoid constructing a weighted mean temperature
scale from mutually inconsistent measurements. Instead, we adopt APOGEE as the internal $T_{\rm eff}$ reference. This
choice is based on the fact that APOGEE has broader and more continuous coverage of effective temperatures and better consistency with PASTEL. In particular, in the high-temperature regime, APOGEE contains stars that remain aligned with the PASTEL scale above $T_{\rm eff} \gtrsim 10000$ K, whereas GALAH has no high-quality stars to provide an overlap anchor at these temperatures and LAMOST LRS shows substantial temperature deviations.

\begin{figure}[ht]
    \centering        \includegraphics[width=\linewidth]        {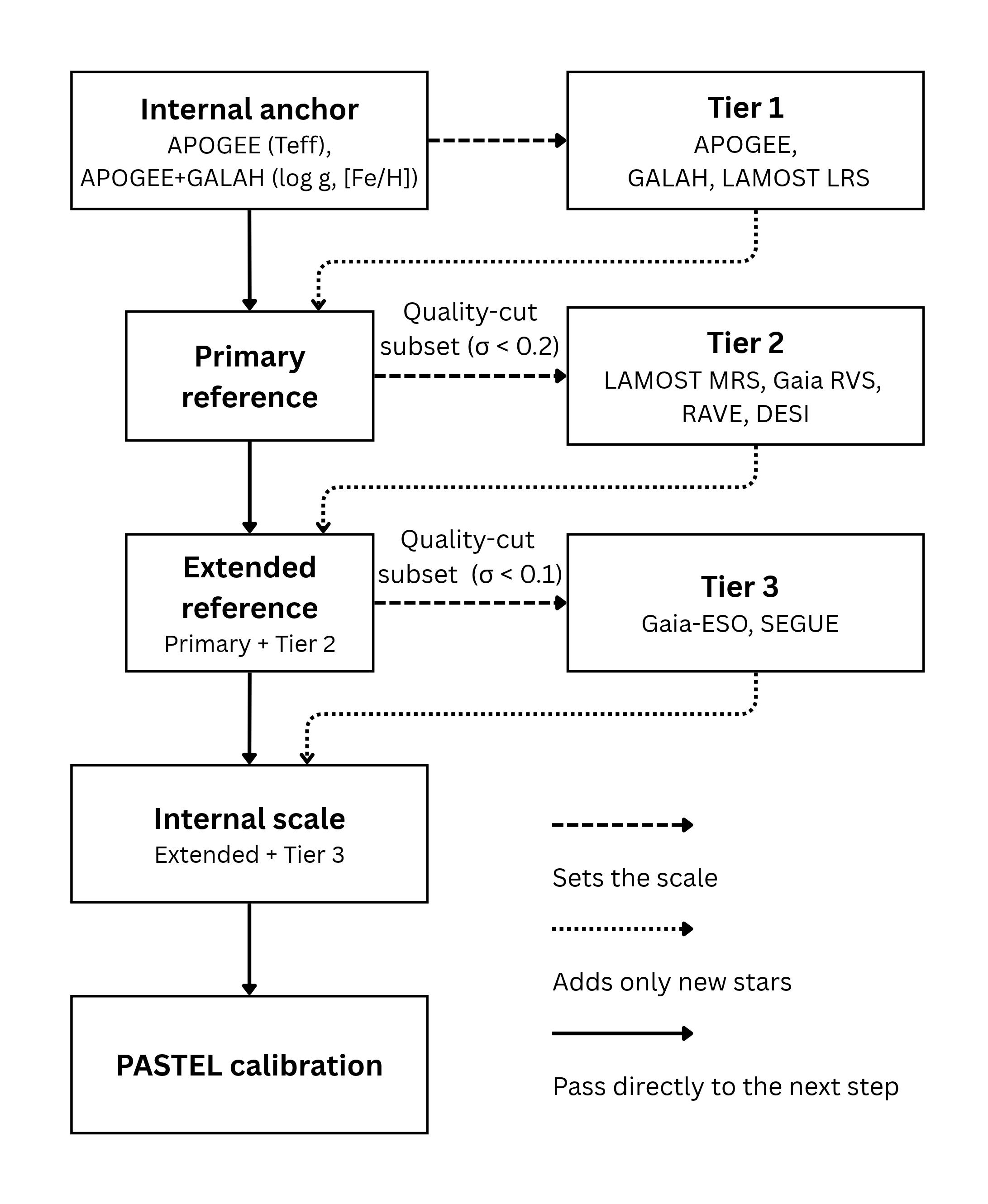}
    \caption{Schematic of the hierarchical calibration. Reference catalogues (left) set the scale for each survey tier (right); calibrated tiers are merged into the next reference, with stars from earlier tiers keeping their values. The internal scale is finally anchored to PASTEL.}
\label{fig:hierarchy}
\end{figure}

On this basis, we adopt the following hierarchical calibration scheme, also summarised schematically in Fig.~\ref{fig:hierarchy}:

(i) For [Fe/H] and $\log g$, we define the initial internal reference scale from the APOGEE--GALAH overlap, using inverse-variance weighted means based on the recalibrated uncertainties (see Sec.~\ref{subsec:uncertainty}).

(ii) LAMOST LRS is then calibrated onto this reference scale for [Fe/H] and $\log g$.

(iii) For $T_{\rm eff}$, APOGEE is adopted as the internal reference scale. GALAH and LAMOST LRS are therefore calibrated independently onto APOGEE.

(iv) The corrected APOGEE, GALAH, and LAMOST LRS measurements are merged to form the primary spectroscopic reference catalogue.

(v) LAMOST MRS, \textit{Gaia} RVS, DESI, and RAVE are calibrated onto this primary reference catalogue. Their calibrated measurements are then added to it to form the extended spectroscopic reference catalogue.

(vi) Finally, SEGUE and \textit{Gaia}-ESO are calibrated onto the extended reference catalogue, which includes the primary reference together with the calibrated LAMOST MRS, \textit{Gaia} RVS, DESI, and RAVE measurements, and are then appended in the final extension step.

At each tier, the newly added surveys are corrected onto the reference scale defined by the previous tiers. The merged catalogue is then updated hierarchically: stars already present in a higher-priority tier keep their adopted parameter set, while newly added surveys contribute only stars with no previous adopted measurement. If a new star is present in more than one survey within the same tier, we adopt the full corrected parameter set from the survey with the smallest combined uncertainty metric,
$U=(\sigma_{T_{\rm eff}}/T_{\rm eff})^2+\sigma_{\log g}^2+\sigma_{\mathrm{[Fe/H]}}^2$.
Here, the $\sigma$ values are the uncertainties obtained after propagating the input errors through the correction procedure. A lower value of $U$ therefore corresponds to a smaller overall uncertainty in the adopted atmospheric parameters. We do not compute a weighted average of multiple calibrated measurements because once survey values have been mapped onto a common scale through data-driven calibration, they are no longer independent estimates, and their uncertainties become partially correlated. A further inverse-variance combination would therefore introduce artificial precision (see e.g.\ \citealt{Dixon2023}).

\subsection{Uncertainty recalibration strategy}
\label{subsec:uncertainty}

The uncertainties reported by different surveys are not necessarily on the same
scale. This can affect both the construction of the reference scale and the
interpretation of the final catalogue uncertainties. We therefore apply an
additional empirical re-calibration of the reported errors. In brief, we use stars observed more than once in each survey to compare the observed
scatter between repeated measurements with the scatter expected from the quoted
uncertainties. The correction is estimated locally in the
$T_{\rm eff}$--$\log g$ plane, so that possible changes of the error behaviour
across parameter space are taken into account. The resulting scale factor is
then used to rescale the input uncertainties before they are propagated through
the calibration procedure.

The full description of this uncertainty recalibration, including the definition of
the normalised residuals and the estimation of the correction factor, is given
in Appendix~\ref{app:error_recalibration}. We apply this recalibration only to the first-tier surveys, where the uncertainties are used directly in the weighted construction of the internal reference scale. For the later tiers, the uncertainty information is heterogeneous across surveys, and the final errors are dominated by the uncertainty propagated through the survey-to-reference correction; we therefore do not apply here an additional uniform recalibration of the input errors.

\subsection{Training the calibration model}
\begin{figure*}[h!]
    \centering        \includegraphics[width=0.9\linewidth]        {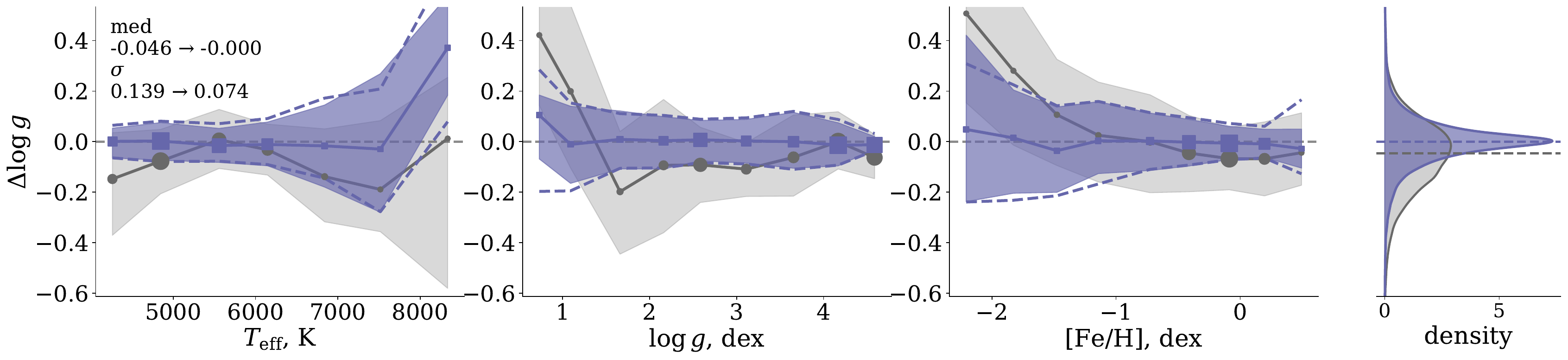}

    \caption{Residuals in LAMOST $\log g$ with respect to the APOGEE+GALAH reference scale, shown as a function of different stellar parameters, before correction (grey) and after correction (purple). Marker size scales with the number of stars in each bin. The shaded region indicates the central 68\% interval of the residuals. The dashed lines show the upper and lower predicted correction bounds.}
    \label{fig:lam_residuals}
\end{figure*}
For each survey and for each atmospheric parameter separately, we train a regression model on the overlap with the adopted reference scale. The target is the parameter correction required to bring the survey measurement onto that reference. We split the overlap sample into training, validation (calibration) and test subsets, and use XGBoost \citep{Chen2016} as a flexible non-linear regressor to model the correction as a function of all three stellar parameters - $T_{\rm eff}$, $\log g$, and [Fe/H]. Using quantile regression, the model predicts not only the median correction but also the lower and upper bounds of the correction distribution. For each parameter, we therefore estimate the 16th, 50th, and 84th percentiles of the correction as functions of the input atmospheric parameters:
\begin{equation*}
\delta X_{p16},\, \delta X_{p50},\, \delta X_{p84}
=
f(T_{\rm eff}, \log g, [\mathrm{Fe/H}]) ,
\end{equation*}
where $X$ denotes $T_{\rm eff}$, $\log g$, or $[\mathrm{Fe/H}]$.

The model setup is described in Appendix~~\ref{app:model_params}; for the training-sample
sizes and fitted hyperparameters, see Table~\ref{tab:hyperparams}.
The uncertainty intervals produced by the XGBoost quantile models are not guaranteed to contain the true values in, for example, 68\% of cases for nominal 68\% intervals. We therefore apply conformalized quantile regression, which combines quantile regression with conformal prediction \citep{Vovk2005,Romano2019}, using the MAPIE implementation \citep[Model Agnostic Prediction Interval Estimator;][]{Cordier2023}. In practice, MAPIE uses a held-out calibration sample to check how often the reference values fall inside the predicted intervals. If the intervals are too narrow, their bounds are expanded by an empirical correction until the calibration-sample coverage is close to the target value. In this work, the target coverage is 68\%, corresponding to the 16th--84th percentile range. This step is performed separately for each survey and for each atmospheric parameter, and modifies only the uncertainty intervals, while the central corrected values are left unchanged. The adjustment is global for each survey--parameter pair: it is estimated over the full validation sample rather than independently in different regions of parameter space. This choice avoids unstable local corrections in sparsely populated regimes, where too few validation stars are available to recalibrate the interval bounds reliably. As a result, the final intervals should be interpreted as globally calibrated predictive intervals; their local reliability still depends on how well a given region of parameter space is represented in the training and validation data.
Fig.~\ref{fig:lam_residuals} shows an example of how the LAMOST LRS $\log g$ residuals relative to the APOGEE/GALAH scale vary across stellar-parameter space, before and after applying the correction. Marker size indicates the relative number of stars available for training in each bin. Bins with fewer stars generally provide weaker constraints on the correction, because the model has fewer examples from which to learn the local relation between surveys. However, a poor correction is not necessarily caused only by a small training sample: it can also reflect larger scatter in the data, which make the relation intrinsically harder to model. This is particularly relevant above $T_{\rm eff}\sim7500$ K, where the training sample becomes sparse and the predicted intervals should therefore be interpreted with caution. We keep these stars in the full catalogue, while the recommended
high-quality selection (see Sec.\ref{sec:catalogue}) is restricted to cooler stars.

\subsection{Applying corrections and building the merged catalogue}

For each survey and parameter, the trained model is applied to the full catalogue to obtain the predicted correction quantiles for every star. The corrected nominal value is defined using the median predicted correction and the lower and upper bounds of the corrected estimate are obtained analogously from the calibrated 16th and 84th percentile corrections,

\begin{equation*}
X_{\rm cal} = X + \delta X_{p50} ,\qquad
X_{\rm low} = X + \delta X_{p16}, \qquad
X_{\rm high} = X + \delta X_{p84}.
\end{equation*}
where $X$ denotes $T_{\rm eff}$, $\log g$, or $[\mathrm{Fe/H}]$.

The calibration correction depends not only on the calibration model but also
on the uncertainties of the input stellar parameters. We therefore propagate
both sources of uncertainty using a Monte Carlo procedure. For each star, we generate $M=300$ Monte Carlo realizations of the input
parameter vector $(T_{\rm eff}, \log g, {\rm [Fe/H]})$, sampling from the
survey-reported uncertainties or, for the first-tier surveys, the recalibrated
uncertainties. Each realization is passed through the calibrated
quantile-regression model, which predicts the 16th, 50th and 84th percentiles
($q_{16}$, $q_{50}$ and $q_{84}$) of the calibration correction.

For each realization, these three quantiles are used to construct a split-normal
(asymmetric Gaussian) approximation to the predictive distribution of the
correction. The distribution is centred on the median prediction, with left and
right standard deviations

\begin{equation*}
\sigma_{\rm left}=q_{50}-q_{16}, \qquad
\sigma_{\rm right}=q_{84}-q_{50}.
\end{equation*}

A single correction is then sampled from this predictive distribution. Applying
the sampled correction to the corresponding sampled survey value yields one Monte Carlo realization of the calibrated parameter. Repeating this
procedure for all $M$ realizations produces the final distribution of calibrated
values, from which the adopted calibrated parameter and its uncertainty interval
are taken as the 50th, 16th and 84th percentiles.

\begin{figure}[ht]
    \centering
        \includegraphics[width=0.65\linewidth]
        {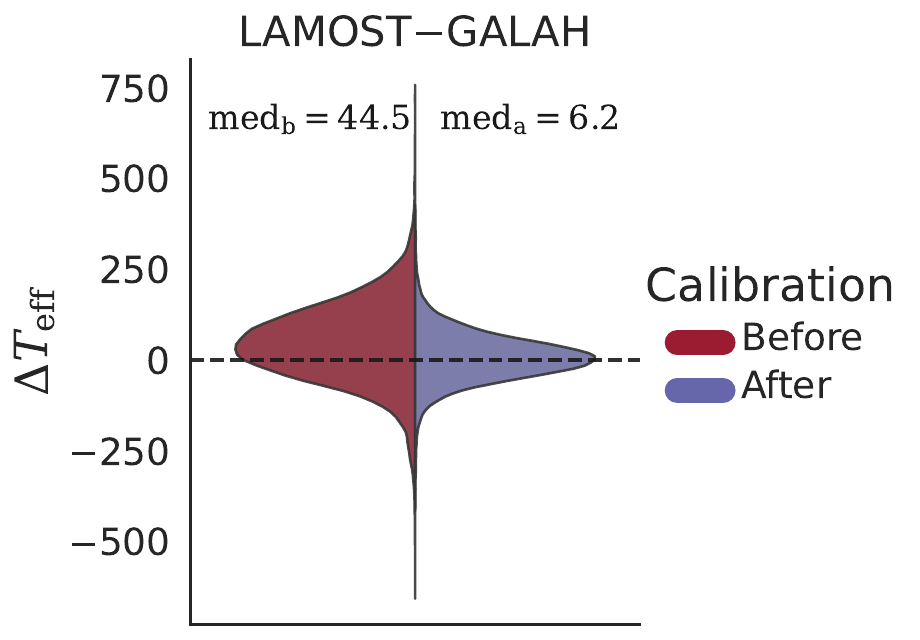}
        \caption{Violin distributions of $\Delta T_{\rm eff}$
(LAMOST$-$GALAH) for stars outside the training sample, before and
after calibration. The dashed line marks zero, and the annotations show
the median offsets before calibration, $\mathrm{med}_{b}$, and after calibration, $\mathrm{med}_{a}$.}
    \label{fig:violin_galah_lamost}
\end{figure}

Because GALAH and LAMOST LRS $T_{\rm eff}$ are calibrated independently onto the APOGEE scale, we verify their mutual consistency using stars outside the APOGEE training sample. After calibration, the median $\Delta T_{\rm eff}$ is reduced from \(44.5\) K to \(6.2\) K, and the distribution becomes substantially narrower (Fig.~\ref{fig:violin_galah_lamost}).
After calibration, the measurements are merged on a star-by-star basis following the hierarchical scheme defined in Sec.~\ref{sec:method}. For [Fe/H] and \(\log g\), we retain the APOGEE--GALAH weighted combination where both surveys are available, and otherwise adopt the available APOGEE or GALAH value before using LAMOST LRS calibrated values. For \(T_{\rm eff}\), where APOGEE defines the reference scale, we retain the available calibrated estimate with the smallest recalibrated uncertainty when multiple measurements exist outside of the APOGEE coverage.

\subsection{Adding other surveys to the scale}
\label{subsec:add_surveys}

Before extending the calibration to additional surveys, we restrict the APOGEE--GALAH--LAMOST reference catalogue to a high-quality subset. The purpose of this step is to prevent poorly constrained measurements from propagating through the next stage of the hierarchy and degrading the effective reference scale. At the same time, constraints applied here remain sufficiently permissive as to have sufficient amount of stars in less populated regions. We require \(\sigma_{\log g}<0.2\) dex, \(\sigma_{\rm [Fe/H]}<0.2\) dex, and \(\sigma_{T_{\rm eff}}/T_{\rm eff}<0.1\). When asymmetric intervals are available, we define \(\sigma=(p84-p16)/2\); otherwise, we use the reported symmetric standard deviation. 

We then extend the scale to the next set of large surveys: RAVE, DESI, LAMOST MRS and \textit{Gaia} RVS. These surveys are calibrated onto the restricted high-quality reference catalogue. For each survey and each atmospheric parameter, we apply the same correction workflow described in Sec.~\ref{sec:method}: the model is trained on the survey overlap with the reference scale, the predictive intervals are calibrated with MAPIE, and the resulting corrections are then applied to the full survey catalogue. After this, we attach new values to the scale from LAMOST MRS, RAVE,
DESI and \textit{Gaia} RVS as described in Sec.~\ref{sec:method}. 

Finally, we apply the same procedure to \textit{Gaia}-ESO and SEGUE, treating them as an additional catalogue-extension tier after \textit{Gaia} RVS, LAMOST MRS, RAVE and DESI. This ordering reflects their limited role in the calibration: \textit{Gaia}-ESO has only a small overlap with the Primary reference sample ($\leq 1000$ stars), while SEGUE has lower spectral resolution and a substantially reduced sample after quality filtering (see Table \ref{tab:quality_filters}). To limit the propagation of residual calibration errors from the
second tier, the training sample for Gaia-ESO and SEGUE is drawn
from the extended reference catalogue after applying stricter
quality cuts than those used for the primary reference:
$\sigma_{\log g}<0.1$\,dex, $\sigma_{\rm [Fe/H]}<0.1$\,dex, and
$\sigma_{T_{\rm eff}}/T_{\rm eff}<0.05$. This retains most
first-tier stars while restricting the second-tier contribution
to well-calibrated measurements.

\subsection{Internal validation}

\begin{figure*}[h!]
    \centering
        \includegraphics[width=0.95\linewidth]
        {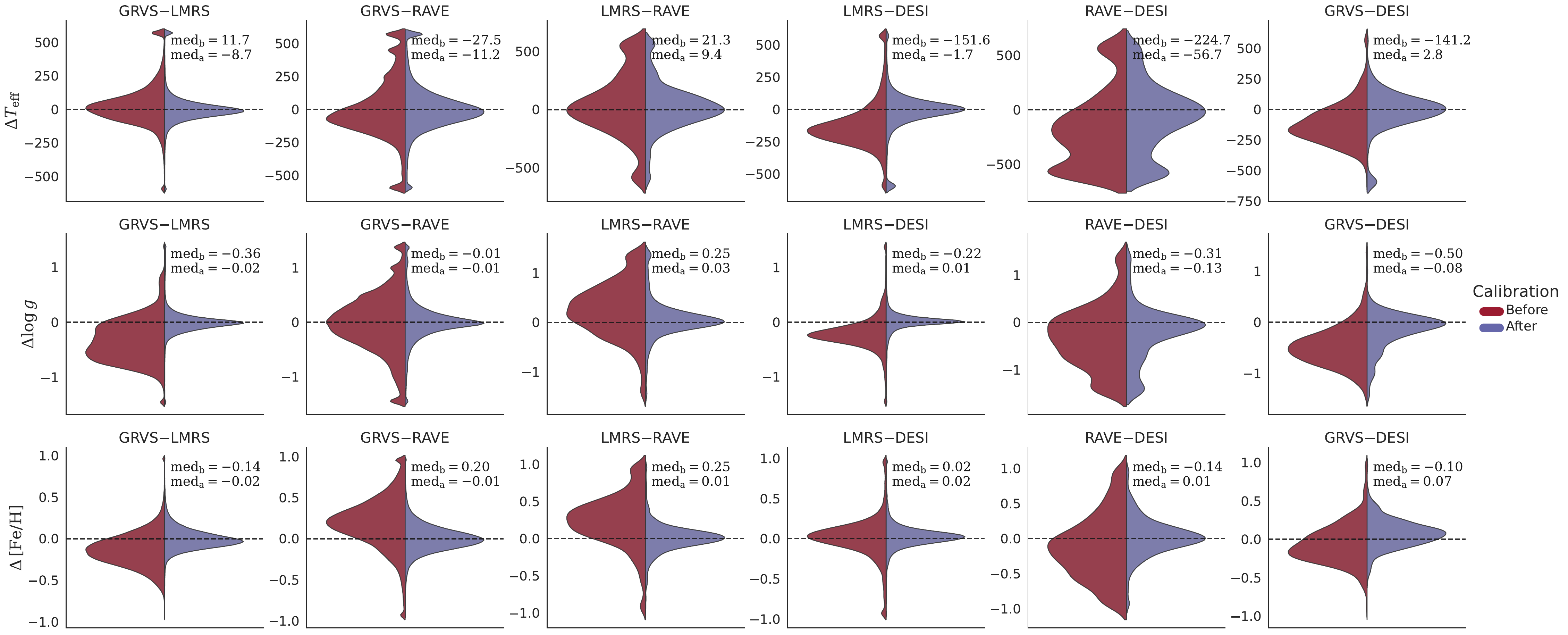}
    \caption{Distributions of survey-to-survey parameter differences between \textit{Gaia} RVS – LAMOST MRS (GRVS - LMRS), \textit{Gaia} RVS – RAVE (GRVS - RAVE), LAMOST MRS – RAVE (LMRS - RAVE), LAMOST MRS - DESI (LMRS - DESI), RAVE - DESI and \textit{Gaia} RVS – DESI (GRVS - DESI) for effective temperature, surface gravity, and metallicity. Split violin plots compare the values before (dark red) and after (purple) calibration. The dashed line marks zero offset. Calibration consistently tightens the distributions and shifts their centres toward zero, indicating improved agreement between surveys across all parameters.}
    \label{fig:violin_surveys}
\end{figure*}

\begin{figure*}[h!]
    \centering
        \includegraphics[width=0.55\linewidth]
        {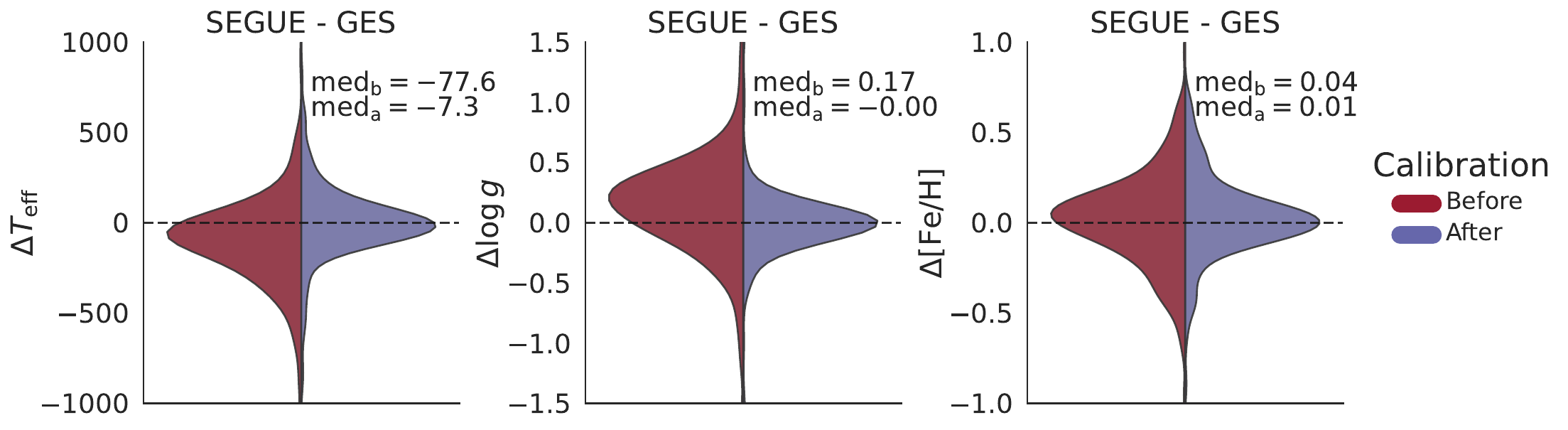}
    \caption{The same as Fig.~\ref{fig:violin_surveys}, but for the third-tier surveys Gaia-ESO and SEGUE.}
    \label{fig:violin_surveys_v2}
\end{figure*}

We assess the internal calibration after the second stage, that is,
after adding \textit{Gaia} RVS, LAMOST MRS, RAVE, and DESI. For each
dataset, we exclude stars that were used during calibration, either for
model training or testing. Using the remaining stars, we construct pairwise
survey intersections and compare the parameter differences before and after
calibration. The resulting distributions are shown in Fig.~\ref{fig:violin_surveys}.
The calibration generally brings the independent survey overlaps closer to
a common scale: the median offsets, shown above each panel, are reduced for
all survey pairs and parameters, and the distributions often become
narrower. The largest changes are seen in comparisons involving DESI,
because these pairs also show some of the largest initial offsets. However, DESI remains less aligned with the other surveys after
calibration than the other survey pairs. Residual differences remain, especially in DESI--RAVE
comparisons and in some $T_{\rm eff}$ and $\log g$ distributions. This
shows that the calibration removes the dominant survey-to-survey trends,
but does not fully erase all survey-specific structures. 
The corresponding comparison for the third-tier surveys is shown in Fig.~\ref{fig:violin_surveys_v2}. Despite the smaller overlap between Gaia-ESO and SEGUE, we are able to see that the calibration brings the distributions of the parameter differences closer to zero and generally makes them narrower. The improvement is particularly clear for $T_{\rm eff}$ and $\log g$, while the [Fe/H] distribution is already relatively well centred before calibration and changes more modestly.

Additional internal-validation diagnostics are provided in Appendix~\ref{app:add_internal_validation}: Appendix~\ref{app:kiel_diagrams} shows the Kiel diagrams before and after calibration, while Appendix~\ref{app:errors} discusses the normalised residual distributions used to assess the calibrated uncertainties. Appendix~\ref{app:snr} examines residual trends with S/N for the second- and third-tier surveys.

\section{External calibration and validation}
\label{sec:validation}

As a final step, we calibrate the internally homogenised parameters against
the PASTEL catalogue, which provides an independent high-resolution
spectroscopic reference scale. PASTEL provides a large compilation of high-resolution spectroscopic $T_{\rm eff}$, $\log g$, and [Fe/H] measurements across the HR diagram. Because it combines results from heterogeneous literature sources, we first construct a consolidated reference sample. We exclude systematically discrepant sources, combine repeated measurements of the same star using inverse-variance weighting, and discard stars with inconsistent measurements. The resulting sample contains $\sim26\,000$
unique stars, of which 14\,360 are in common with SoS. The external calibration is
applied after the survey-to-survey homogenisation and is restricted to
a linear function of the stellar parameters, so that it removes remaining
global trends in $T_{\rm eff}$, $\log g$, and [Fe/H] without
fine-tuning to match PASTEL exactly. The
corrected values are reported in the final catalogue (see
Tab.~\ref{tab:catalog_columns}) as teff\_cal, logg\_cal, and feh\_cal, and are
the recommended parameters for scientific use. The details of 
reference-sample construction, calibration equations, and uncertainty 
propagation are described in Appendix~\ref{app:pastel_calibration}.

\subsection{Validation on \textit{Gaia} benchmark stars}
We validate the homogenised parameters against the \textit{Gaia} FGK benchmark stars catalogue \citep[GBS]{Soubiran2024}. These stars provide an external validation scale, with fundamental $T_{\rm eff}$ and $\log g$ determined independently of spectroscopy from angular diameters, bolometric fluxes, parallaxes, and stellar evolution constraints. The third version of the catalogue contains 192 benchmark stars, while parameters are provided for 201 entries including a small number of stars with indirect angular diameter estimates. The metallicities in the catalogue were adopted from the literature sources, mainly from PASTEL, this should be considered when assessing the agreement in [Fe/H].

\begin{figure*}
    \centering    \includegraphics[width=0.9\textwidth]{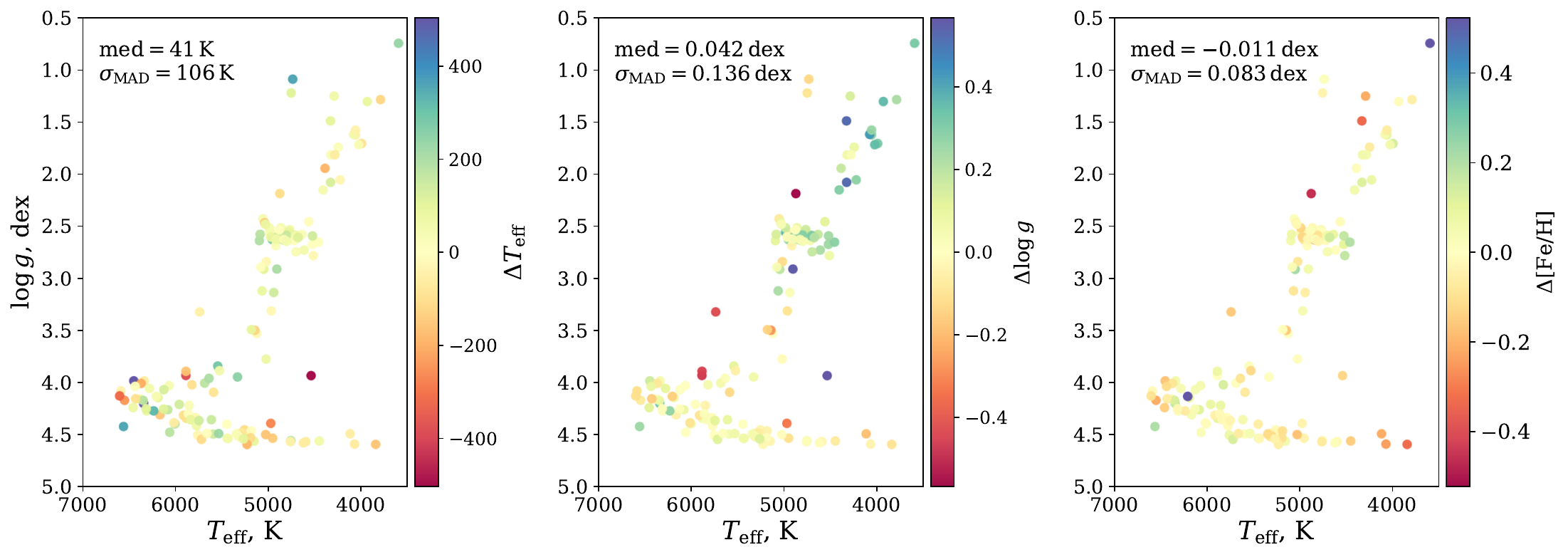}
    \caption{Comparison with GBS in the Kiel diagram. Stars matched between the SoS DR3 catalogue and the benchmark sample are colour-coded by the parameter residuals $\Delta T_{\rm eff}$, $\Delta \log g$, and $\Delta [\mathrm{Fe/H}]$, where $\Delta X = X^{\mathrm{SoS}} - X^{\mathrm{GBS}}$.}
    \label{fig:gbs_kiel}
\end{figure*}

After cross-matching the benchmark sample with our final catalogue, we examined the residuals between the parameters derived in this work and the benchmark values across the Kiel diagram. In particular, we produced three Kiel diagrams, colour-coded by $\Delta T_{\rm eff}$, $\Delta \log g$ and $\Delta [\mathrm{Fe/H}]$ (Fig.~\ref{fig:gbs_kiel}). The median residual and scatter are reported in each panel. Overall, the agreement is good for most benchmark stars, with only a few objects showing large discrepancies. In a catalogue of this size, the calibration is intended to improve the global parameter scale rather than to force agreement for every individual benchmark star, so a small number of outliers is expected.

\subsection[Validation of log g with APOKASC]{Validation of $\log g$ with APOKASC}

We further assessed the quality of the homogenised surface gravities using the latest APOKASC catalogue, APOKASC-3 \citep{Pinsonneault2025}, which combines APOGEE spectroscopy with Kepler asteroseismology for evolved stars in the Kepler fields. This catalogue provides asteroseismic surface gravities for a large sample of red giants and therefore offers an external reference particularly well suited for validating $\log g$ in the giant regime.

\begin{figure*}
    \centering
    \includegraphics[width=0.9\textwidth]{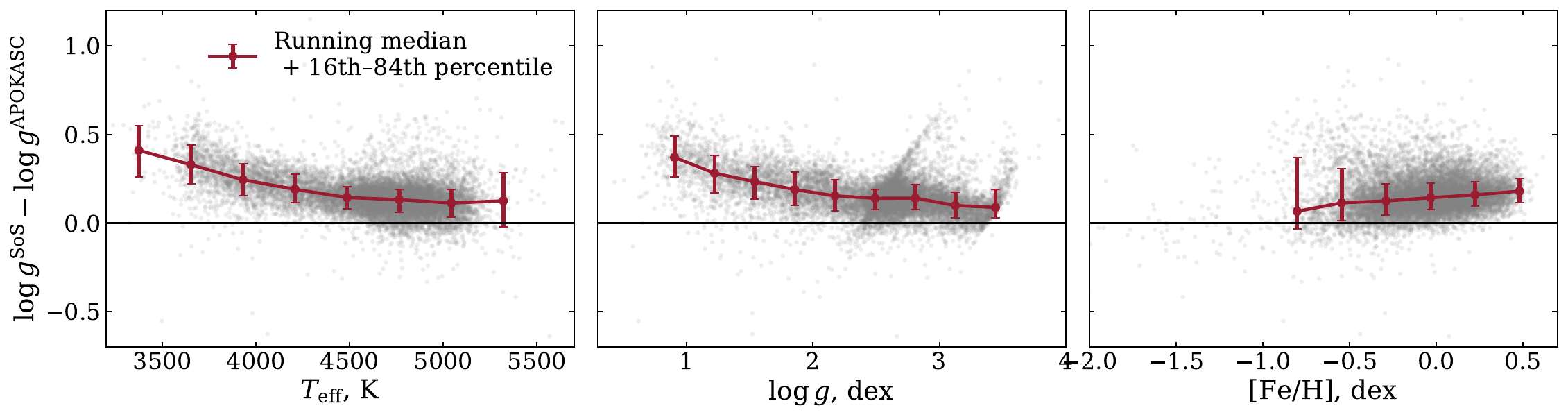}

    \caption{Residuals in surface gravity with respect to APOKASC-3 as a function of $T_{\rm eff}$, $\log g$ and $[\mathrm{Fe/H}]$. In each panel, the solid curve shows the median trend of $\Delta \log g = \log g^{\mathrm{SoS}} - \log g^{\mathrm{APOKASC}}$, while the error bars indicate the corresponding spread.}
    \label{fig:apokasc_logg_trends}
\end{figure*}

We analyse the residuals,
$\Delta\log g = \log g^{\rm SoS} - \log g^{\rm APOKASC}$,
as a function of the stellar parameters using running medians and the
16th--84th percentile ranges, as shown in
Fig.~\ref{fig:apokasc_logg_trends}. The comparison shows that the SoS gravities are systematically higher than
the APOKASC-3 asteroseismic gravities. We find a median residual of
$\Delta\log g = 0.146$ dex, with a $\sigma_{MAD}$ of $0.078$ dex. This
offset is not unexpected, because the internal SoS scale is tied to the
APOGEE/GALAH spectroscopic reference scale, and APOGEE surface gravities
are known to show similar residual systematic differences with respect to
asteroseismic gravities. \citet{Masseron2017} showed that these discrepancies
depend on evolutionary state, with distinct behaviour for red-giant-branch
and red-clump stars. Despite the calibration efforts implemented in APOGEE
DR17, differences between the APOGEE spectroscopic scale and the APOKASC-3
asteroseismic scale remain. The APOGEE DR19 \citep{Meszaros2025} validation likewise identifies residual
trends with stellar parameters and evolutionary state and applies a new
calibration based on APOKASC-3. We do not apply an
additional asteroseismic recalibration in the present work; consequently, the
SoS $\log g$ values remain more closely tied to the spectroscopic
APOGEE-based scale than to the APOKASC asteroseismic scale.

\subsection{Validation of metallicity on star clusters}

\begin{figure*}
    \centering
\includegraphics[width=0.85\textwidth]{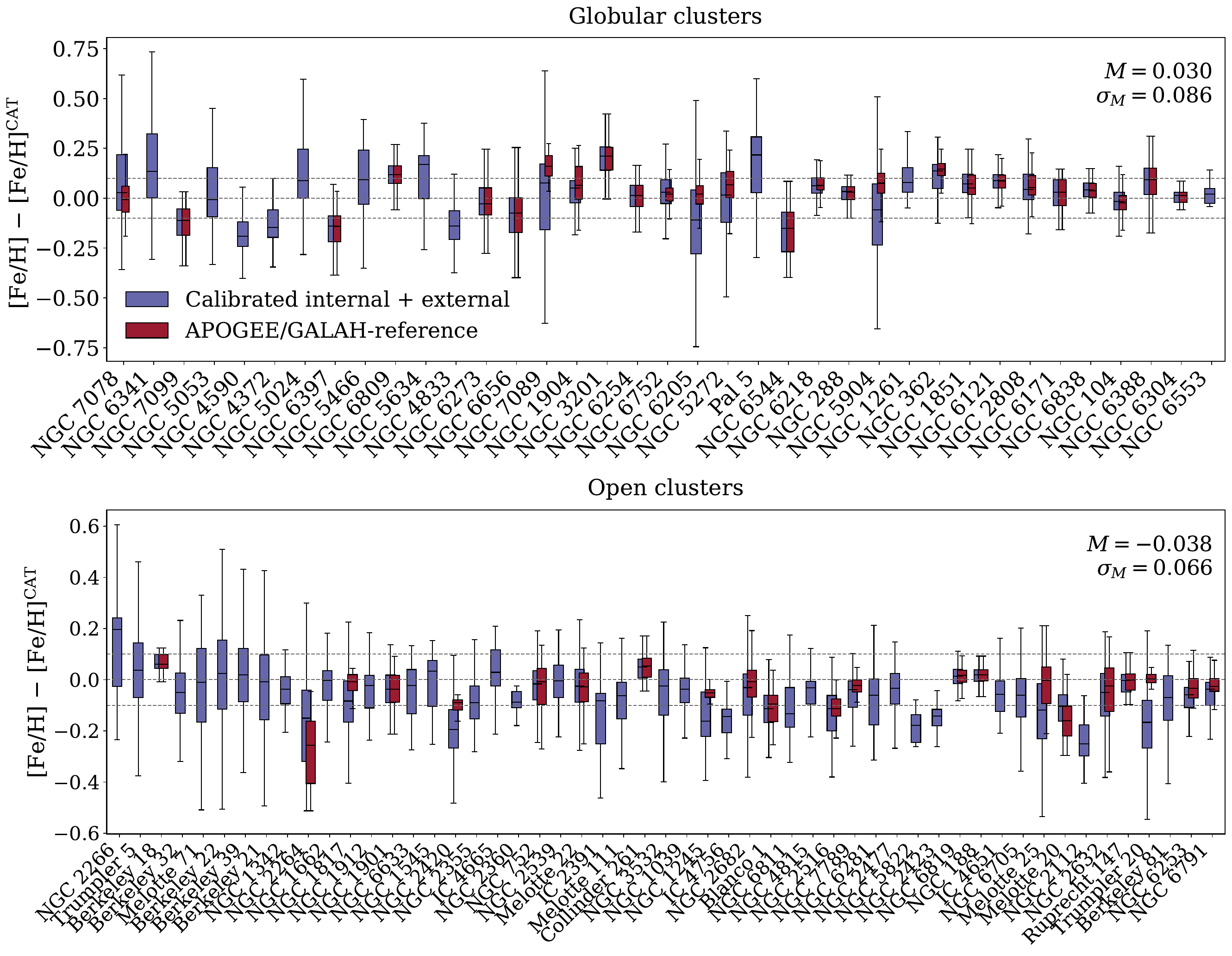}
    \caption{Comparison of calibrated metallicities with literature reference values for globular and open clusters. Residuals are shown for the different calibration paths discussed in the text. Each box spans the 25th to 75th percentiles, the central line marks the median, and the whiskers extend to the most extreme points within 1.5 times the interquartile range.}
    \label{fig:clusters}
\end{figure*}

To assess the robustness of the final metallicity scale, we compare the catalogue metallicities with literature values for globular and open clusters, as shown in Fig.~\ref{fig:clusters}. The catalogue metallicities used in this comparison correspond to the final calibrated values, after both steps of our procedure: first, the individual survey measurements are homogenised onto the internal APOGEE--GALAH reference scale; second, the resulting scale is adjusted externally using PASTEL. As a reference, we also show the original APOGEE and GALAH metallicities after applying only their direct external calibration to the PASTEL scale. For this comparison, we exclude stars with $T_{\rm eff} > 7500$ K, because such hot stars are not covered by a reliable reference in our internal calibration scheme. Apart from this temperature cut, all available cluster stars are used.

For globular clusters, reference metallicities were adopted from \citet{Harris1996, Harris2010}, and probable members were selected using \citet{Vasiliev2021} with a membership probability threshold of $p>0.8$. For open clusters, high-quality literature metallicities were taken from \citet{Netopil2016}, and cluster members were selected following \citet{Hunt2024}. For globular clusters, most of the stars in our final validation sample come from the original APOGEE and GALAH surveys. Since this regime is already in good agreement with PASTEL, the applied corrections are generally small. Nevertheless, in the relatively few cases where the final metallicities originate from other surveys, we still find good agreement with the reference scale. The small systematic offset of $0.030 \pm 0.086$\,dex indicates that the calibrated metallicities remain consistent with the external literature values from \citet{Harris1996, Harris2010}.

For open clusters, we find a systematic offset of approximately
$-0.038 \pm 0.066$\,dex relative to the metallicity scale of \citet{Netopil2016}.
This offset is seen for different surveys and calibration paths, suggesting
that it is unlikely to be caused only by the internal homogenisation procedure. A similar tendency was reported in the previous SoS release \citep[see their Fig.~8]{Turchi2025}, where the same type of cluster-based comparison was performed for SoS and other catalogues with machine-learning-derived stellar parameters. In that comparison, the SoS metallicities showed relatively small offsets, compatible with those found for the original APOGEE metallicities, whereas the other machine learning-based catalogues showed offsets of the same sign but with larger amplitudes. Therefore, the offset seen here should not be interpreted as a specific weakness of the present catalogue. Rather, it likely reflects a residual difference between homogeneous spectroscopic metallicity scales and the heterogeneous literature cluster scale compiled by \citet{Netopil2016}. Given its modest amplitude, we do not apply an additional correction based on
this comparison.

\section{Catalogue description}
\label{sec:catalogue}

The SoS DR3 catalogue provides homogenised atmospheric parameters together with their lower and upper
uncertainty bounds, the source survey of each adopted parameter, and two
recommended quality flags for 10.13 million stars compiled from nine large spectroscopic surveys. The parameters are placed on a common internal scale defined by APOGEE and GALAH, and are subsequently externally calibrated to the PASTEL reference scale. Survey-dependent systematics are corrected using data-driven regression models. The description of columns is provided in Table~\ref{tab:catalog_columns}.

\begin{table*}
\caption{Description of the main catalogue columns.}
\label{tab:catalog_columns}
\centering
\setlength{\tabcolsep}{4pt}
\renewcommand{\arraystretch}{1.10}

\begin{tabular}{@{}p{0.32\textwidth}p{0.10\textwidth}p{0.52\textwidth}@{}}
\hline
\hline
Column & Unit & Description \\
\hline

\multicolumn{3}{@{}l}{\emph{Identifiers and coordinates}} \\
\texttt{sos\_id} & -- & Unique SoS identifier. \\
\texttt{gaia\_dr3\_source\_id} & -- & \textit{Gaia} DR3 source identifier, where available. \\
\texttt{ra}, \texttt{dec} & deg & Adopted ICRS coordinates. \\
\texttt{param\_source} & -- & Source survey of the adopted atmospheric-parameter set. \\

\hline
\multicolumn{3}{@{}l}{\emph{Quality flags}} \\
\texttt{high\_quality} & -- & Default high-quality subset recommended for general scientific use. \\
\texttt{feh\_quality} & -- & Metallicity-optimised subset. \\

\hline
\multicolumn{3}{@{}l}{\emph{Homogenised and externally calibrated parameters}} \\
\texttt{teff}, \texttt{logg}, \texttt{feh} &
K, dex, dex &
Homogenised internal-scale $T_{\rm eff}$, $\log g$, and [Fe/H]. \\

\texttt{teff\_lo/hi}, \texttt{logg\_lo/hi}, \texttt{feh\_lo/hi} &
K, dex, dex &
16th and 84th percentile bounds of the internal-scale parameters. \\

\texttt{teff\_cal}, \texttt{logg\_cal}, \texttt{feh\_cal} &
K, dex, dex &
Final externally calibrated $T_{\rm eff}$, $\log g$, and [Fe/H]. \\

\texttt{teff\_cal\_lo/hi}, \texttt{logg\_cal\_lo/hi},
\texttt{feh\_cal\_lo/hi} &
K, dex, dex &
16th and 84th percentile bounds of the externally calibrated parameters. \\

\hline
\end{tabular}

\tablefoot{The suffixes \texttt{\_lo} and \texttt{\_hi} denote the 16th and
84th percentile bounds, respectively. For stars without input uncertainties (mainly from APOGEE and GALAH), calibrated values are provided without uncertainty bounds. Externally calibrated values require all three internal-scale
parameters and are therefore not provided for stars with incomplete
parameter sets.}
\end{table*}

We provide two quality flags for the homogenised catalogue. The first flag,
\texttt{high\_quality}, defines the default high-quality subset recommended
for general scientific use. The second flag, \texttt{feh\_quality}, defines
a stricter subset for studies where the accuracy of [Fe/H] is the main
requirement. The \texttt{feh\_quality} sample is included within the
\texttt{high\_quality} sample, but it should not be treated as a better
selection for all stellar parameters.

The \texttt{high\_quality} flag is based on the final calibrated
predictive uncertainties of the corrected parameters and on independent
\textit{Gaia} source-level diagnostics. For the spectroscopic parameters, we require
\[
\sigma_{\log g} < 0.2~\mathrm{dex}, \qquad
\sigma_{[\mathrm{Fe/H}]} < 0.2~\mathrm{dex}, \qquad
\sigma_{T_{\rm eff}}/T_{\rm eff} < 0.05,
\]
where the uncertainties are derived from the calibrated 16th--84th percentile
prediction intervals of the quantile-regression model. We additionally require
$T_{\rm eff}<7500$~K.

To reduce the impact of problematic source-level measurements, the same flag
also requires $\mathrm{RUWE}<1.4$, $\texttt{ipd\_frac\_multi\_peak}\leq10$, no
match to the \textit{Gaia} QSO or galaxy candidate lists, no \textit{Gaia} photometric variability
flag, and $\texttt{non\_single\_star}=0$. These criteria are not used in the
spectroscopic calibration itself, but provide an independent check on source
reliability, blending, variability, multiplicity, and non-stellar contamination.
The resulting high-quality sample contains
6.10 million stars, corresponding to
60.2\% of the catalogue.

The metallicity-focused subset, marked by
\texttt{feh\_quality} flag, applies a stricter uncertainty
selection, requiring $\sigma_{\log g} < 0.1~\mathrm{dex}$ and
$\sigma_{[\mathrm{Fe/H}]} < 0.1~\mathrm{dex}$
together with the recommended-quality \textit{Gaia} filters and the same temperature
limit. This selection improves the external validation performance in
$[\mathrm{Fe/H}]$, but does not produce a comparable improvement in
$T_{\rm eff}$ or $\log g$. We therefore use it only as a metallicity-optimised
flag rather than as the default high-quality catalogue selection. The
metallicity-optimised subset contains 3.69 million stars,
corresponding to 36.5\% of the catalogue. The validation performance of both selections is summarized in
Table~\ref{tab:quality_external_summary}. 
While the quality selections reduce the dispersion, the median
offsets drift slightly away from zero. For globular clusters, the
metallicity-optimised subset consists almost entirely of APOGEE and
GALAH stars, for which the only correction applied is the external
calibration on PASTEL; this alone reduces the median offset from 0.063 (the result before calibration on PASTEL)
to 0.046\,dex.

\begin{table}
\caption{External validation summary for the recommended subsets.}
\label{tab:quality_external_summary}
\centering
\setlength{\tabcolsep}{2.5pt}
\begin{tabular}{@{}p{0.27\columnwidth}p{0.19\columnwidth}cc@{}}
\hline
\hline
Sample & Parameter & High-quality & [Fe/H]-opt. \\
\hline
GBS
& $T_{\rm eff}$, K
& 30 / 107
& -- \\

*
& $\log g$, dex
& 0.033 / 0.125
& -- \\

*
& [Fe/H], dex
& $-$0.016 / 0.083
& 0.003 / 0.066 \\

APOKASC
& $\log g$, dex
& 0.142 / 0.076
& -- \\

Globular clusters
& [Fe/H], dex
& 0.039 / 0.072
& 0.046 / 0.060 \\

Open clusters
& [Fe/H], dex
& $-$0.035 / 0.054
& $-$0.018 / 0.042 \\
\hline
\end{tabular}
\tablefoot{Each entry gives median $\Delta$ / $\sigma_{\rm MAD}$. For the
stellar benchmark samples, $\Delta = X_{\rm SoS} - X_{\rm ref}$. For
clusters, the values are computed from the cluster-level metallicity
residuals. An asterisk means that the validation sample is the same as
in the row above. }
\end{table}

\section{Science application: the metal-poor tail of thin-disc-like stars}
\label{sec:sci_case}

As a simple illustration of the scientific use of the homogenised catalogue, we revisit the population of metal-poor stars on disc-like orbits. Such stars are of particular interest because they may preserve signatures of the earliest phases of Milky Way disc formation. Previous studies \citep{FernandezAlvar2024,Bellazzini2024,Nepal2024,Zhang2024, GonzalezRivera2024} have shown that fast-rotating, near-planar stars exist well below the canonical thin-disc metallicity limit of [Fe/H] $\sim -0.7$ dex. 

For this science-case analysis, we construct a working sample following the astrometric quality selection adopted by \citet{FernandezAlvar2024}. Specifically, we require $\mathrm{RUWE}<1.4$, positive parallaxes, and relative parallax uncertainties below 20 percent. These criteria limit the inclusion of sources with less reliable astrometric solutions or poorly constrained distances.

At the same time, we adopt broader quality cuts on the calibrated atmospheric
parameters than in the high-quality subset described above. This choice is
motivated by the design of the calibration procedure: the predicted and
calibrated uncertainties reflect the local coverage of the training sample and
therefore tend to increase in sparsely populated regions of parameter space,
especially at low metallicity. A very restrictive uncertainty cut would thus
preferentially remove metal-poor stars and suppress the population that this
science case aims to test. We therefore require
\begin{equation*}
    \sigma_{[\mathrm{Fe/H}]} < 0.3~\mathrm{dex}, \qquad
    \sigma_{\log g} < 0.3~\mathrm{dex}, \qquad
    \sigma_{T_{\rm eff}}/T_{\rm eff} < 0.1 ,
\end{equation*}
together with $T_{\rm eff}<7500$\,K.

We then derive the stellar velocities and orbital actions following the general
procedure of \citet[][their Sect.~3.1]{Kordopatis2023}. We use \textit{Gaia} DR3 proper
motions and RVs, together with the photogeometric distances from
\citet{BailerJones2021}, to reconstruct the six-dimensional phase-space
coordinates of the stars. The orbit calculations are performed with
\texttt{galpy} \citep{Bovy2015}. 
The orbital actions are computed in the \texttt{MWPotential2014} Milky Way
potential using the Staeckel approximation \citep{Binney2012, Sanders2016} implemented in \texttt{galpy}. From
these quantities, we obtain $J_r$, $J_z$, $L_z$, $V_\phi$, and $Z_{\max}$, and
define
\begin{equation*}
    J_{\rm tot}=J_r+|L_z|+J_z .
\end{equation*}

We select metal-poor thin-disc-like stars following the criteria adopted by
\citet{FernandezAlvar2024}. The dynamical selection is based on the normalised
orbital actions and on the maximum height above the Galactic plane, $L_z/J_{\rm tot} > 0.95, \left|(J_z-J_r)/J_{\rm tot}\right| < 0.05, Z_{\max} < 1.5~{\rm kpc}.$
These cuts select stars on cold, disc-like orbits with dominant azimuthal
motion and limited vertical excursions. We then restrict the sample to the
metal-poor regime by requiring $[\mathrm{Fe/H}]_{\rm corr} < -0.55 $.

\begin{figure}
    \centering
    \includegraphics[width=0.8\columnwidth]{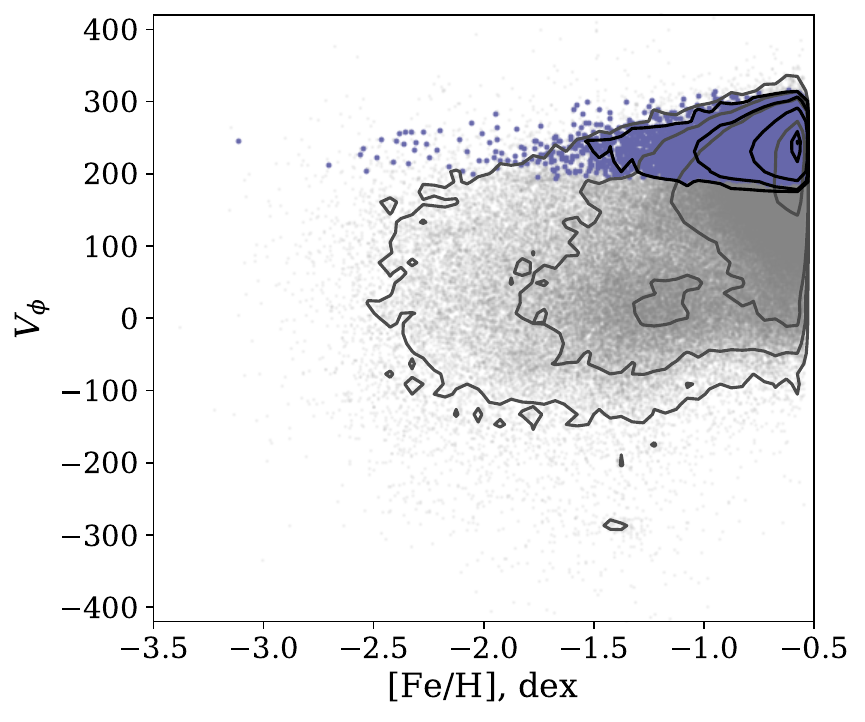}
    \caption{Distribution of stars in the $V_\phi$--[Fe/H] plane. The parent sample is shown in grey, while the thin-disc-like subsample selected from the homogenised catalogue is overplotted in violet, with contours highlighting the highest-density regions.}
    \label{fig:vphi_feh_tail}
\end{figure}

Figure~\ref{fig:vphi_feh_tail} shows the distribution of the parent sample in the $V_\phi$--[Fe/H] plane, with the thin-disc-like subsample over-plotted in violet colour. The selected stars form a rapidly rotating sequence that extends well below the usual thin-disc metallicity range. In particular, the metal-poor tail remains clearly populated down to
$\mathrm{[Fe/H]} \sim -1.5$, with 29 stars below
$\mathrm{[Fe/H]} = -2$. We therefore confirm the presence of a
metal-poor population on disc-like orbits using a broader, internally
consistent multi-survey parameter scale. The selected sample is not
identical to those reported previously, because some Gaia RVS stars are
removed by our quality cuts, while additional candidates are contributed
by LAMOST, GALAH, and APOGEE.

\section{Conclusions}
\label{sec:conclusion}

We have constructed a homogenised catalogue of stellar atmospheric parameters from nine major spectroscopic surveys using a hierarchical calibration framework. The internal scale is defined from APOGEE, GALAH, and LAMOST LRS, with APOGEE adopted as the reference for $T_{\rm eff}$ and the APOGEE--GALAH overlap used to define the reference for $\log g$ and [Fe/H]. LAMOST MRS, \textit{Gaia} RVS, RAVE, DESI, \textit{Gaia}-ESO, and SEGUE were then calibrated onto this scale, and the final parameters were externally corrected using the PASTEL compilation. The resulting catalogue provides homogenised $T_{\rm eff}$, $\log g$, and [Fe/H] for 10.13 million stars.

The calibration includes an empirical treatment of survey-dependent uncertainties. For APOGEE and LAMOST LRS, the reported uncertainties were recalibrated using repeated observations, while GALAH uncertainties were placed on the same scale through comparisons with the corrected surveys. Survey-to-reference corrections were modelled with XGBoost quantile regression, the predictive intervals were calibrated with MAPIE, and the input uncertainties were propagated through the correction procedure with Monte Carlo sampling. The catalogue therefore provides both corrected parameters and asymmetric uncertainty intervals.

Internal validation shows that the calibration reduces the main survey-to-survey offsets and improves the agreement between independent overlap samples. Some survey-specific structure remains, especially in comparisons involving DESI and RAVE, showing that the calibration places the surveys on a common scale but does not remove all limitations of the original parameter spaces. External validation confirms that the final scale is broadly consistent with independent reference data: \textit{Gaia} benchmark stars show good agreement across the Kiel diagram, the metallicity scale is supported by globular and open cluster comparisons, and APOKASC-3 highlights a remaining positive offset in $\log g$ for red giants relative to the asteroseismic scale.

We provide two recommended quality selections. The \texttt{high\_quality} flag defines the default subset for general scientific use and contains 6.10 million stars, while the stricter \texttt{feh\_quality} flag is optimised for metallicity-focused applications and contains 3.69 million stars. As an example application, we used the catalogue to confirm the presence of a metal-poor, rapidly rotating population on
thin-disc-like orbits, demonstrating that the homogenised multi-survey
scale can be used to study rare Galactic populations with improved statistics and consistent stellar parameters.

\section{Data availability}
The SoS DR3 final catalogue and the individual calibrated survey catalogues are available electronically on the SoS portal at the Space Science Data Center\footnote{\url{https://gaiaportal.ssdc.asi.it/SoS}} and at CDS\footnote{\url{https://vizier.cds.unistra.fr/viz-bin/VizieR}}. 

\begin{acknowledgements}
{\bf People.} 
We thank our colleagues for helpful discussions: F.~Gran, P.~A.~Palicio, G.~Thomas. We thank the anonymous referee for the constructive comments that helped improve this manuscript. 

{\bf Funding.} 
Funded by the European Union (ERC-2022-AdG, {\em "StarDance: the non-canonical evolution of stars in clusters"}, Grant Agreement 101093572, PI: E. Pancino). Views and opinions expressed are however those of the author(s) only and do not necessarily reflect those of the European Union or the European Research Council. Neither the European Union nor the granting authority can be held responsible for them. EL acknowledges support from the ERC Consolidator Grant funding scheme (project ASTEROCHRONOMETRY, \url{https://www.asterochronometry.eu}, G.A. n. 772293).
We are grateful for the fruitful discussions during the team meetings
at the International Space Science Institute in Bern. This work was
supported by the ISSI International Team project
``AsteroSHOP: large Spectroscopic surveys HOmogenisation Program''
(PI: G. Thomas).
{\bf Software.} This research made use of \texttt{Astropy}, \texttt{NumPy}, \texttt{pandas}, \texttt{SciPy}, \texttt{scikit-learn}, \texttt{XGBoost}, \texttt{MAPIE}, \texttt{Optuna}, \texttt{Matplotlib}, and \texttt{galpy}. This research has made use of the SIMBAD database and the VizieR catalogue access tool, both operated at CDS, Strasbourg, France. Preliminary data exploration relied heavily on TopCat. 
\end{acknowledgements}

\bibliographystyle{bibtex/aa.bst}   
\bibliography{references_cleaned}

\appendix

\section{Error recalibration strategy}
\label{app:error_recalibration}

Before combining measurements from different surveys, we need to make their
reported uncertainties statistically consistent. APOGEE and LAMOST LRS include
repeated spectroscopic observations for some stars. These repeated observations
allow us to test the internal precision of their uncertainties directly. In
contrast, the final GALAH catalogue provides only one set of stellar parameters
per object, obtained after internally co-adding the spectra of the same star.
For this reason, GALAH uncertainties cannot be tested in the same way. We
therefore recalibrate the APOGEE and LAMOST LRS uncertainties using duplicate
observations, while GALAH is treated separately through its consistency with the
already corrected uncertainties from the other surveys.

If the measurement errors are unbiased and Gaussian, the differences between two
repeated measurements, normalised by the quadrature sum of their reported
uncertainties, should follow a standard normal distribution ($\mu = 0$, $\sigma = 1$). In practice, this
condition is often not satisfied. The normalised distribution can be either too narrow or too broad, indicating that the reported uncertainties are overestimated or underestimated, respectively (see eq.~A.1). We use this behaviour to estimate a scaling factor for the reported uncertainties and to bring them onto a more consistent error scale.

We first remove objects flagged as \texttt{non\_single\_star} and stars with \textit{Gaia} DR3 $\mathrm{RUWE}>1.4$, to reduce possible contamination from stars in multiple systems or RV variable stars. The repeated spectroscopic observations are then grouped by \textit{Gaia} DR3 \texttt{source\_id}. For each stellar parameter ($T_{\rm eff}$, $\log g$, and [Fe/H]), we compute all pairwise differences within each group and normalize them by the quadrature sum of the corresponding reported uncertainties:

\begin{equation}
\Delta x_{\rm norm} =
\frac{x_i-x_j}{\sqrt{\sigma_i^2+\sigma_j^2}} .
\end{equation}

where $x_i$ and $x_j$ are the measurements and $\sigma_i$, $\sigma_j$ are corresponding errors. If the reported uncertainties are accurate, the distribution of $\Delta x_{\rm norm}$ should be a normal distribution centred at zero with unit variance. Deviations from this behaviour indicate local under- or overestimation of the formal errors. 

Since the error scale changes across the parameter space, we divide the
sample into bins in the $T_{\rm eff}$--$\log g$ plane. This allows us to
group stars with similar error behaviour. In each bin, we estimate a
correction factor, $\kappa$, which rescales the reported uncertainties. The
factor $\kappa$ is chosen so that the scatter of $\Delta x_{\rm norm}$
becomes close to one, as expected for correctly estimated Gaussian errors. We compute $\kappa$ only in bins where the distribution of
$\Delta x_{\rm norm}$ is close enough to Gaussian. If the distribution is
strongly non-Gaussian, its standard deviation is no longer a reliable
estimate of the uncertainty scale. To identify such bins, we use two
criteria. The first is a non-Gaussianity index,
\begin{equation}
{\rm NGI} =
\sqrt{
{\rm skew}^2 +
\left(\frac{{\rm kurtosis}}{3} - 1\right)^2
},
\end{equation}
where skewness and kurtosis are computed in the standard way. The second
criterion compares three estimates of the scatter: the standard deviation
from a Gaussian fit, $\sigma$, the scaled median absolute deviation,
$\sigma_{\rm MAD}$, and the interquartile-range estimate,
$\sigma_{\rm IQR}$:

\begin{equation}
\delta =
|\sigma-\sigma_{\rm MAD}| +
|\sigma_{\rm MAD}-\sigma_{\rm IQR}| +
|\sigma-\sigma_{\rm IQR}|.
\end{equation}

\begin{figure}[ht]
    \centering
        \includegraphics[width=0.85\linewidth]        {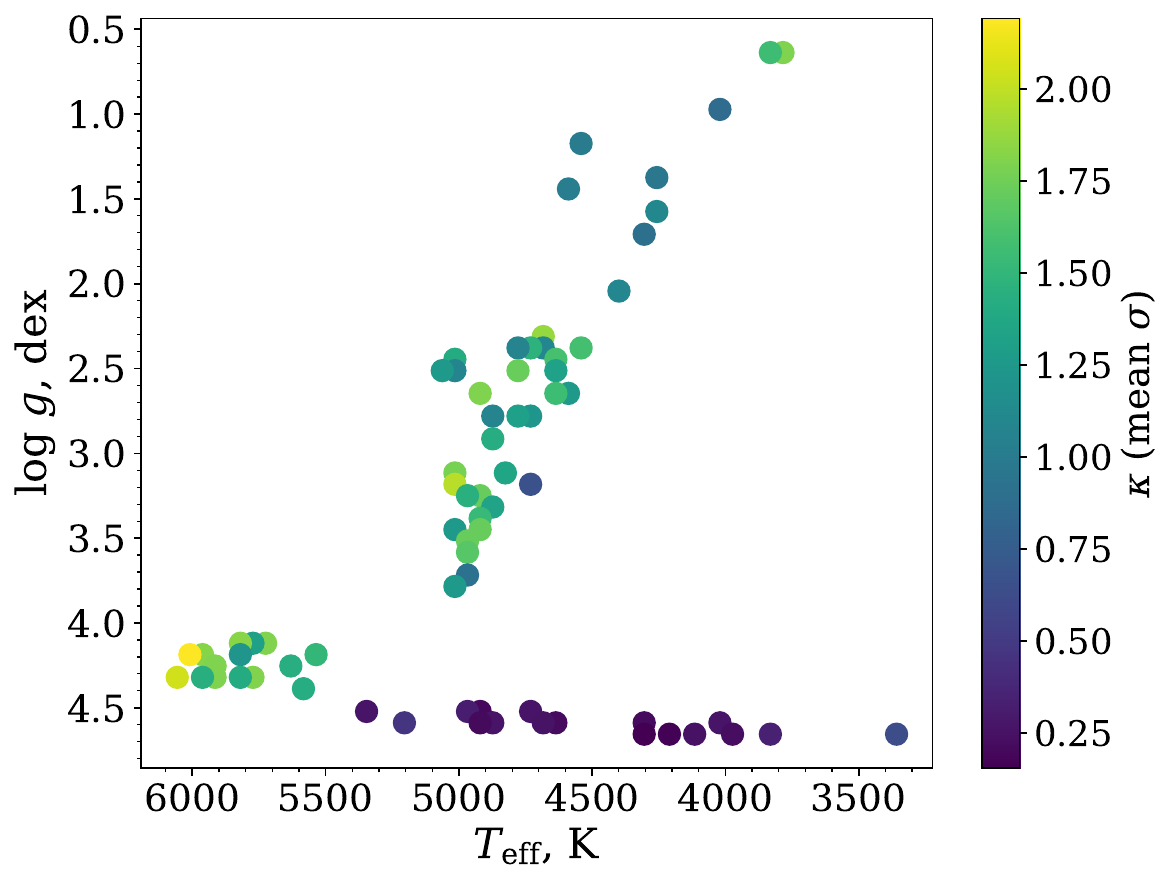}
    \caption{Robust error normalisation factor $\kappa$ for $\log g$ in APOGEE as a function of effective temperature and surface gravity. The colour scale shows $\kappa$, defined as the mean of the standard deviation, MAD- and IQR-based scatter estimates.}

    \label{fig:logg_error_map}
\end{figure}

We retain only bins with ${\rm NGI}<0.3$ and $\delta<0.35$. These limits
were chosen from Monte Carlo tests as a compromise between robustness and
sample retention. For each accepted bin, we define $\kappa$ as the average
of the three scatter estimates,
\begin{equation}
\kappa = \frac{1}{3}
\left(\sigma + \sigma_{\rm MAD} + \sigma_{\rm IQR}\right).
\end{equation}
The values of $\kappa$ measured in the accepted bins are then used to train
a K-nearest-neighbours regressor \citep{Cover1967}, which predicts
$\kappa$ continuously across the full $T_{\rm eff}$--$\log g$ plane. The
reported formal uncertainties are then rescaled by the predicted value of
$\kappa$. Figure~\ref{fig:logg_error_map} shows the resulting $\kappa$ map for APOGEE $\log g$ as an example. The APOGEE $\log g$ uncertainties are overestimated for cool
dwarfs, while they are underestimated for giants, subgiants, and hotter dwarfs.

As it was discussed before, for GALAH we cannot apply the same approach as the multiple spectra of one object are co-added to obtain the final result. To place GALAH uncertainties on the same scale, we compare GALAH measurements with APOGEE and LAMOST after the latter two have been corrected using their $\kappa$ maps. For each matched star, we compute

\begin{equation*}
\Delta x_{\rm norm} =
\frac{x_{\rm GALAH}-x_{\rm ref}}
{\sqrt{\sigma_{\rm GALAH}^2+\sigma_{\rm ref}^2}} .
\end{equation*}

where the reference is either APOGEE or LAMOST with recalibrated uncertainties. We determine in each $T_{\rm eff}$--$\log g$ bin the value of $\kappa$ required to bring the scatter of $\Delta x_{\rm norm}$ to unity.

The $\kappa$ estimates obtained from the GALAH--APOGEE and GALAH--LAMOST overlaps are broadly consistent. As a consistency check, we trained a random forest \citep{Breiman2001} classifier to distinguish $\kappa$ values derived from the GALAH--APOGEE and GALAH--LAMOST overlaps. Its limited discriminating power (accuracy 50--65\%) indicates that the two overlap sets do not support strongly different recalibration maps. We therefore combine them into a single training set and fit a unified KNN model for GALAH across the parameter space.

\section{Calibration models setup}
\label{app:model_params}

Table~\ref{tab:hyperparams} summarizes the training samples
and final hyperparameters used for the calibration models. To ensure an independent evaluation, measurements referring to the same \textit{Gaia} source were always assigned to the same train, validation, or test subset, whenever repeated \textit{Gaia} source identifiers appeared in either the reference catalogue or the survey being calibrated. This prevents information from the same star from entering more than one subset. For samples without repeated \textit{Gaia} identifiers, we used a simple random split.

The listed hyperparameters are the final values adopted for each model tuned with \texttt{Optuna} hyperparameter optimization framework \citep{Akiba2019}.
The number of estimators ($n_{\rm est}$) was fixed manually to reduce the
computational cost of the final hyperparameter search. Preliminary \texttt{Optuna} runs
in which $n_{\rm est}$ was also optimised typically preferred values of about
800--1200 trees, supporting the adopted choice. The remaining
hyperparameters were optimised within predefined search ranges. For example, \texttt{reg\_lambda} was optimised between $10^{-2}$ and $50$ on a
logarithmic scale, such that large values indicate that stronger regularization
was preferred for the corresponding model.

\begin{table*}
\caption{Training samples and hyperparameters of all calibration models.}
\label{tab:hyperparams}
\centering
\small
\setlength{\tabcolsep}{4pt}

\begin{tabular}{lll rrr cccccc}
\hline\hline
Survey &
Parameter &
Reference\tablefootmark{a} &
$N_{\rm train}$ &
$N_{\rm val}$ &
$N_{\rm test}$ &
$n_{\rm est}$ &
$\eta$ &
$d_{\rm max}$ &
$m_{\rm child}$ &
$f_{\rm sub}$ &
$\lambda$ \\
\hline

GALAH &
$T_{\rm eff}$ &
APOGEE &
24\,978 &
8\,358 &
8\,350 &
800 &
0.052 &
5 &
24 &
0.827 &
0.013 \\

LAMOST LRS &
$T_{\rm eff}$ &
APOGEE &
140\,339 &
46\,882 &
46\,175 &
1000 &
0.051 &
5 &
7 &
0.921 &
7.140 \\

LAMOST LRS &
$\log g$ &
APO+GAL &
176\,471 &
58\,588 &
58\,626 &
1000 &
0.058 &
6 &
5 &
0.624 &
0.048 \\

LAMOST LRS &
$[\mathrm{Fe/H}]$ &
APO+GAL &
177\,679 &
58\,304 &
58\,701 &
1000 &
0.072 &
6 &
16 &
0.908 &
0.576 \\

\textit{Gaia} RVS &
$T_{\rm eff}$ &
Primary &
171\,001 &
57\,001 &
57\,001 &
1000 &
0.095 &
10 &
1 &
0.727 &
0.889 \\

\textit{Gaia} RVS &
$\log g$ &
Primary &
171\,001 &
57\,001 &
57\,001 &
1000 &
0.015 &
9 &
15 &
0.516 &
7.272 \\

\textit{Gaia} RVS &
$[\mathrm{Fe/H}]$ &
Primary &
171\,001 &
57\,001 &
57\,001 &
1000 &
0.011 &
10 &
1 &
0.500 &
8.985 \\

LAMOST MRS &
$T_{\rm eff}$ &
Primary &
866\,711 &
290\,700 &
290\,690 &
1000 &
0.005 &
7 &
2 &
0.950 &
6.232 \\

LAMOST MRS &
$\log g$ &
Primary &
866\,711 &
290\,700 &
290\,690 &
1000 &
0.037 &
10 &
18 &
0.746 &
1.377 \\

LAMOST MRS &
$[\mathrm{Fe/H}]$ &
Primary &
866\,711 &
290\,700 &
290\,690 &
1000 &
0.012 &
10 &
2 &
0.850 &
0.030 \\

DESI &
$T_{\rm eff}$ &
Primary &
113\,615 &
37\,901 &
37\,854 &
1000 &
0.156 &
4 &
26 &
0.703 &
0.822 \\

DESI &
$\log g$ &
Primary &
113\,615 &
37\,901 &
37\,854 &
1000 &
0.150 &
5 &
29 &
0.889 &
17.889 \\

DESI &
$[\mathrm{Fe/H}]$ &
Primary &
113\,652 &
37\,796 &
37\,880 &
1000 &
0.067 &
5 &
15 &
0.883 &
20.657 \\

RAVE &
$T_{\rm eff}$ &
Primary &
10\,212 &
3\,404 &
3\,404 &
800 &
0.020 &
6 &
8 &
0.719 &
6.816 \\

RAVE &
$\log g$ &
Primary &
10\,212 &
3\,404 &
3\,404 &
800 &
0.009 &
8 &
9 &
0.583 &
7.670 \\

RAVE &
$[\mathrm{Fe/H}]$ &
Primary &
10\,212 &
3\,404 &
3\,404 &
800 &
0.031 &
4 &
5 &
0.804 &
6.474 \\

\textit{Gaia}-ESO &
$T_{\rm eff}$ &
Extended &
1\,840 &
614  &
613 &
200 &
0.077 &
8 &
4 &
0.838 &
2.416 \\

\textit{Gaia}-ESO &
$\log g$ &
Extended &
1\,840 &
614  &
613  &
200 &
0.099 &
8 &
9 &
0.984 &
0.085 \\

\textit{Gaia}-ESO &
$[\mathrm{Fe/H}]$ &
Extended &
1\,840 &
614 &
613  &
200 &
0.085 &
5 &
8 &
0.592 &
5.896 \\

SEGUE &
$T_{\rm eff}$ &
Extended &
14\,094 &
4\,671 &
4\,685 &
1000 &
0.019 &
5 &
6 &
0.660 &
6.003 \\

SEGUE &
$\log g$ &
Extended &
14\,094 &
4\,671 &
4\,685 &
1000 &
0.028 &
6 &
12 &
0.749 &
0.705 \\

SEGUE &
$[\mathrm{Fe/H}]$ &
Extended &
14\,094 &
4\,671 &
4\,685 &
1000 &
0.034 &
15 &
4 &
0.802 &
1.003 \\

\hline
\end{tabular}

\tablefoot{
For all models, the input features are the original survey estimates of
$T_{\rm eff}$, $\log g$, and $[\mathrm{Fe/H}]$. All models use the XGBoost
quantile-regression objective, \texttt{reg:quantileerror}. The columns list the
number of estimators $n_{\rm est}$, learning rate $\eta$, maximum depth
$d_{\rm max}$, minimum child weight $m_{\rm child}$, subsampling fraction
$f_{\rm sub}$, and regularization parameter $\lambda$.
\tablefoottext{a}{
Reference used as the calibration target for each survey--parameter model.
APO+GAL denotes the inverse-variance weighted APOGEE--GALAH reference used
for $\log g$ and $[\mathrm{Fe/H}]$.
Primary denotes the reference catalogue constructed from APOGEE, GALAH,
and calibrated LAMOST LRS, after the quality cuts described in the Sec.\ref{subsec:add_surveys}.
Extended denotes the primary reference catalogue after adding the calibrated
LAMOST MRS, \textit{Gaia} RVS, RAVE, and DESI datasets, and applying the stricter quality cuts
described in the same section.
}}

\end{table*}

\section{Objects without \textit{Gaia} source identifiers}
\label{app:no_gaia}
After the \textit{Gaia}-ID-based catalogue was constructed, we also add
survey entries without assigned \textit{Gaia} source identifiers. These objects are
not used to construct the calibration overlaps, but they are retained in
the final catalogue. For this part of the catalogue, we apply a separate
coordinate- and parameter-based merging procedure. First, within each
survey, we remove duplicate entries with exactly identical coordinates.
We then search for possible matches between surveys within a radius of
1 arcsec. For each candidate match, we compute a score that combines the
angular separation with the consistency of the available original,
uncalibrated survey quantities:

\begin{equation}
S =
\exp\left[
-\frac{1}{2}
\left(
\frac{\mathrm{sep}^2}{\sigma_{\rm pos}^2}
+
\sum_p
\left(
\frac{\Delta p}{\sigma_p}
\right)^2
\right)
\right],
\end{equation}
where \(\mathrm{sep}\) is the angular separation, \(\sigma_{\rm pos}\)
is the corresponding positional uncertainty, \(\Delta p\) is the
difference in the original survey quantity \(p\), and \(\sigma_p\) is the
corresponding uncertainty. Depending on the available information,
\(p\) denotes either the original atmospheric parameters $T_{\rm eff}$,
$\log g$, and [Fe/H], or the available magnitudes. For each source, we
keep the candidate with the highest score. 

The entries without \textit{Gaia} source identifiers are then merged using the
same survey-priority logic described above (see Sec.~\ref{sec:method}). In summary, LAMOST LRS is adopted first. We do not have APOGEE and GALAH stars left without \textit{Gaia} \texttt{source\_id} after our quality cuts. 
For sources not already included in LAMOST LRS, we choose among
LAMOST MRS, DESI, and RAVE using the combined uncertainty score and
adopt the full calibrated parameter set from the selected survey. SEGUE
is added in the final step for objects not already represented in the
previous layers.

\section{Additional internal-validation diagnostics}
\label{app:add_internal_validation}

\subsection{Kiel diagrams before and after calibration}
\label{app:kiel_diagrams}

\begin{figure*}[h!]
    \centering
        \includegraphics[width=0.9\linewidth]
        {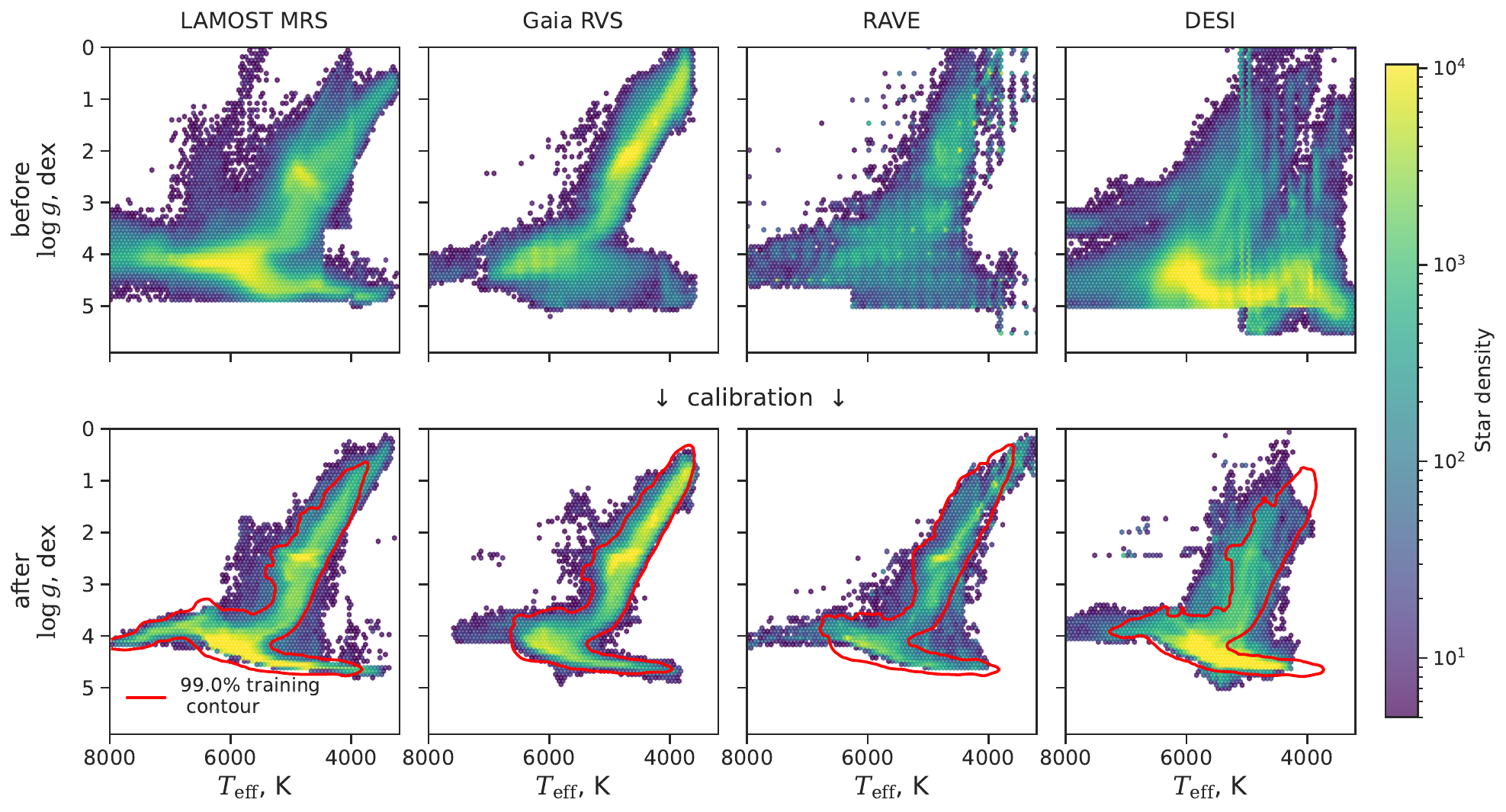}
        \caption{Kiel diagrams for LAMOST MRS, \textit{Gaia} RVS, RAVE and DESI stars before and after calibration to the common reference scale. Colours reflect local stellar density. Post-calibration panels show improved consistency across surveys, with reduced systematic offsets and a more coherent structure of the stellar locus in both temperature and gravity.}
    \label{fig:kiel_corrected}
\end{figure*}

\begin{figure}[h!]
    \centering
        \includegraphics[width=\linewidth]
        {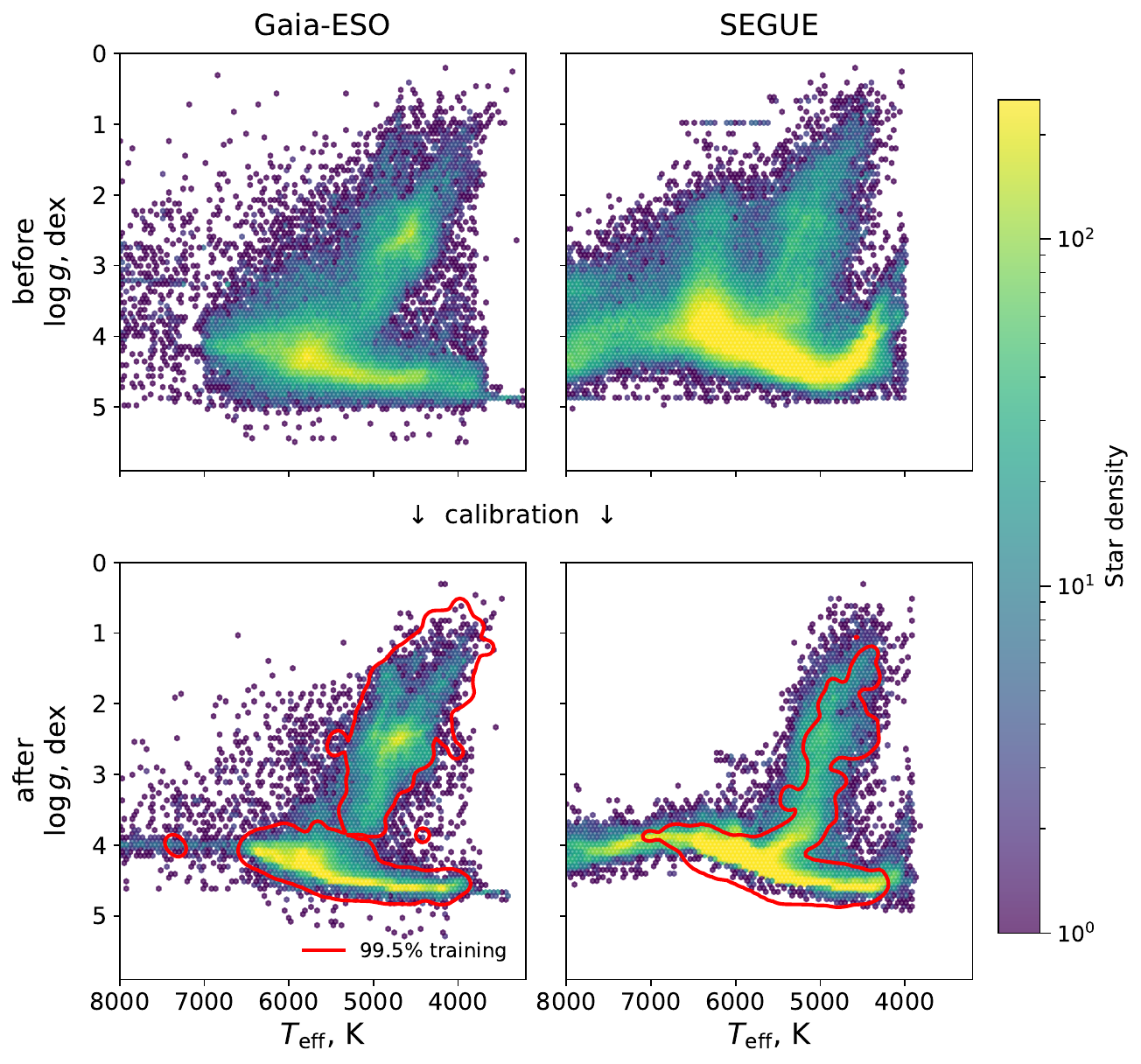}
        \caption{The same as Fig.~\ref{fig:kiel_corrected}, but for the third-tier surveys. Because the training samples are less populated, we show the 99.5\% training sample contour.}
    \label{fig:kiel_corrected_t3}
\end{figure}

For each survey, we present Kiel diagrams before and after calibration to illustrate the improved consistency across the stellar parameter space (Fig.~\ref{fig:kiel_corrected}). After calibration, the key features, such as the red clump and the giant branch, appear
more sharply defined. The improvement is strongest for LAMOST MRS, \textit{Gaia} RVS,
and RAVE. DESI also moves closer to the common reference scale, but its
post-calibration Kiel diagram is not fully aligned with the training-sample
locus, especially for cooler giants. Some residual vertical structures also
remain. This suggests that part of the DESI-specific parameter structure is
not fully removed by the calibration, and that DESI-based parameters should
be used with particular attention to the reported uncertainties and quality
flags in these regions. Fig.~\ref{fig:kiel_corrected_t3} shows the same comparison for the third-tier surveys, Gaia-ESO and SEGUE. Because their training samples cover the parameter space less uniformly, we use a higher enclosed fraction for the contour. The more limited coverage is clearly visible and affects the calibration. Nevertheless, the calibrated distributions still show improvements. Namely, the distribution of hot stars in Gaia-ESO becomes more regular, while the peculiar feature seen among SEGUE dwarfs cooler than $\sim4500$~K becomes less prominent after calibration.

The improved alignment should still be interpreted with caution. The calibration models are trained to reproduce the dominant trends in the overlap samples and apply corrections based on the reported position of each star in parameter space. They are therefore not expected to identify individual problematic or intrinsically peculiar objects. An outlier in the original catalogue may remain an outlier after calibration, even if its deviation is reduced, and a clean post-calibration Kiel diagram does not by itself guarantee that every unusual object has been assigned physically correct parameters. This is a general limitation of data-driven homogenisation: the method is designed to improve the consistency of the atmospheric parameters for the main stellar populations in the catalogue, not to classify or validate all peculiar stars individually.

\subsection{Error distribution after calibration}
\label{app:errors}

We assess globally whether the uncertainties of the calibrated measurements after Stage~2
(calibration of \textit{Gaia} RVS, LAMOST MRS, RAVE and DESI) are internally consistent.
To this end, we analyse the distribution of normalised residuals, defined as

\begin{equation}
z = \frac{\Delta p}{\sqrt{\sigma_1^2 + \sigma_2^2}},
\end{equation}

where $\Delta p$ is the difference between the calibrated parameters from a given survey pair,
and $\sigma_1$, $\sigma_2$ are their corresponding uncertainties.

\begin{figure*}
    \centering
        \includegraphics[width=\linewidth]{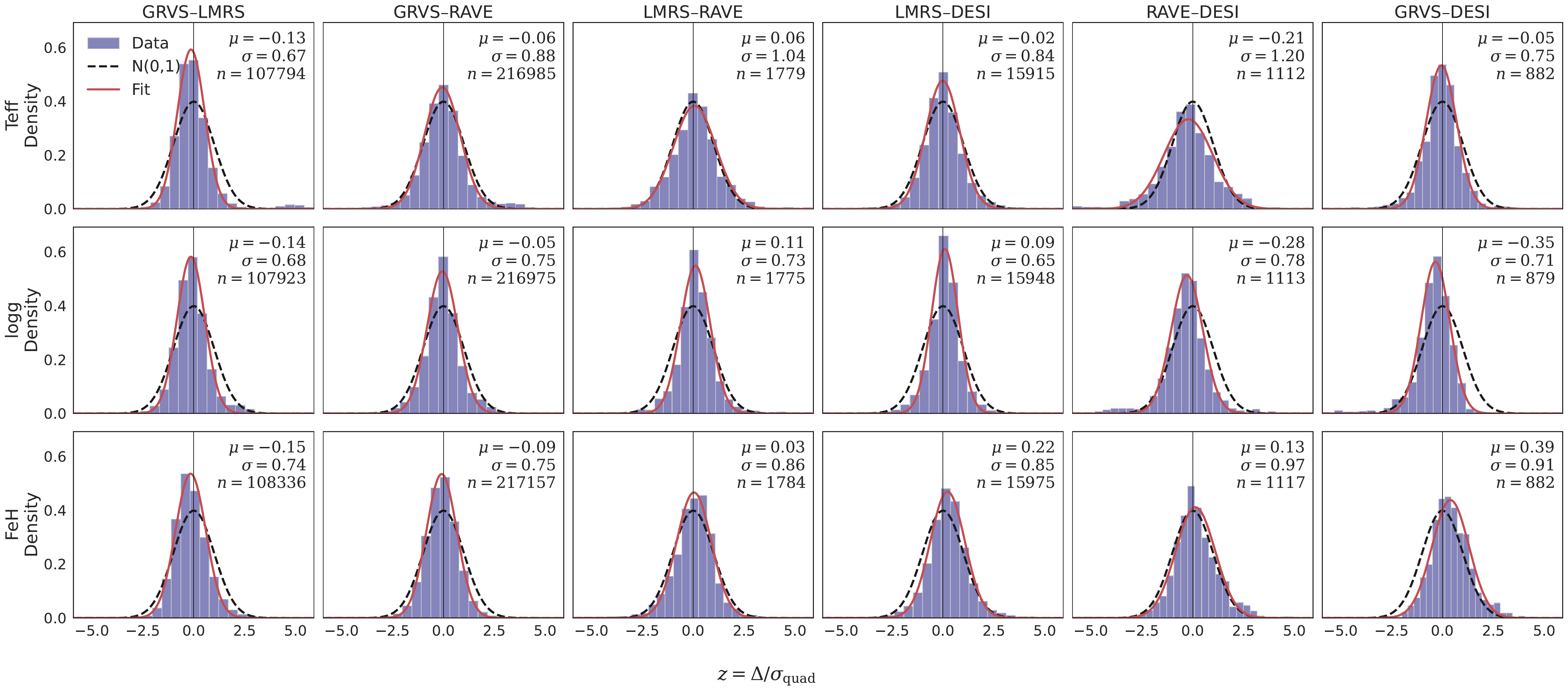}
        \vspace{-0.5em}

    \caption{
    Distributions of normalised residuals ($z$) after Stage~2 calibration for
    $T_{\rm eff}$, $\log g$, and
[Fe/H].
    Rows correspond to the three stellar parameters, while columns correspond to the different survey pairs. The histograms show the distributions of the normalised residuals, compared with a standard normal distribution and with a Gaussian fit to the robust core, defined after 3-$\sigma_{\rm MAD}$ clipping.
    }
    \label{fig:z_all_params}
\end{figure*}

Figure~\ref{fig:z_all_params}
shows the distributions of $z$ for $T_{\mathrm{eff}}$, $\log g$, and [Fe/H], respectively, for all survey pairs. For $T_{\mathrm{eff}}$, we observe that uncertainties associated with \textit{Gaia}~RVS tend to be globally
overestimated, while those from LAMOST~MRS appear to be underestimated.
DESI exhibits a population of outliers, indicating that for a subset of objects the calibration
correction was insufficient. These cases are not fully accounted for by uncertainties estimated
within the procedure adopted here. For $\log g$, the uncertainties across all surveys appear to be systematically overestimated in a
similar manner, with DESI again showing prominent outliers. For [Fe/H], \textit{Gaia}~RVS uncertainties may be overestimated, while those from RAVE and LAMOST~MRS appear
to be broadly consistent. DESI shows both an increased number of outliers and a noticeable skewness
in the $z$ distribution, suggesting that DESI [Fe/H] values should be treated with caution even
after calibration. This, however, is not against the overall improvement achieved through the
calibration process.

These trends indicate that the reported uncertainty intervals should be interpreted with some caution for \textit{Gaia} RVS and LAMOST MRS. We therefore retain these surveys in the catalogue, but one should rely on the provided quality flags and validation diagnostics when selecting high-precision samples.

\subsection{Inspection of trends with signal-to-noise ratio}
\label{app:snr}

\begin{figure*}
    \centering
        \includegraphics[width=0.8\linewidth]{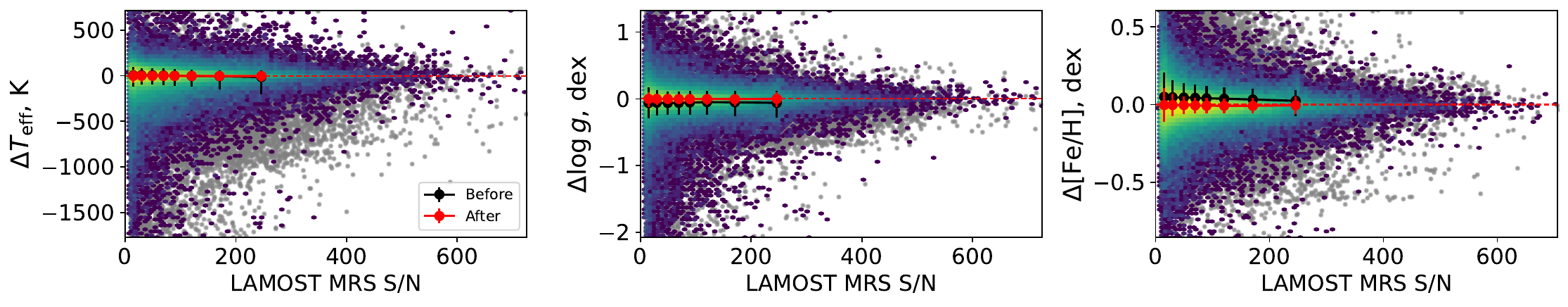} \\
        \includegraphics[width=0.8\linewidth]{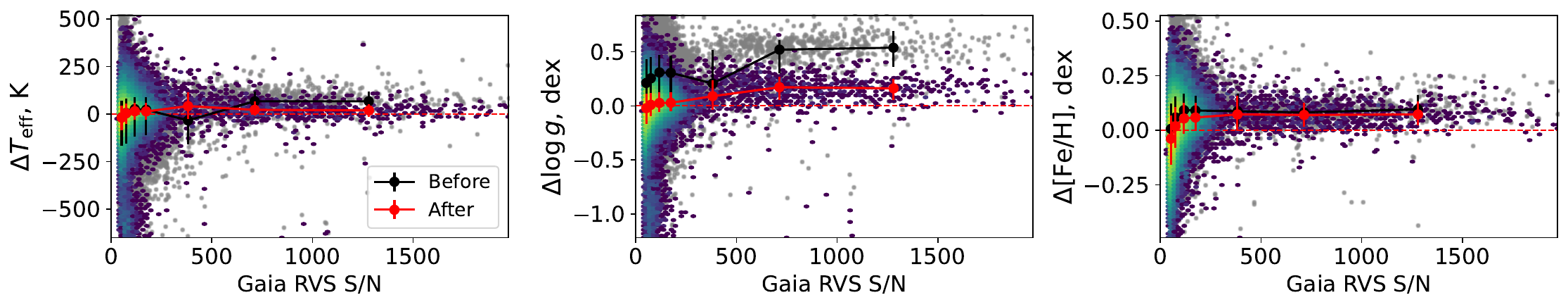} \\
        \includegraphics[width=0.8\linewidth]{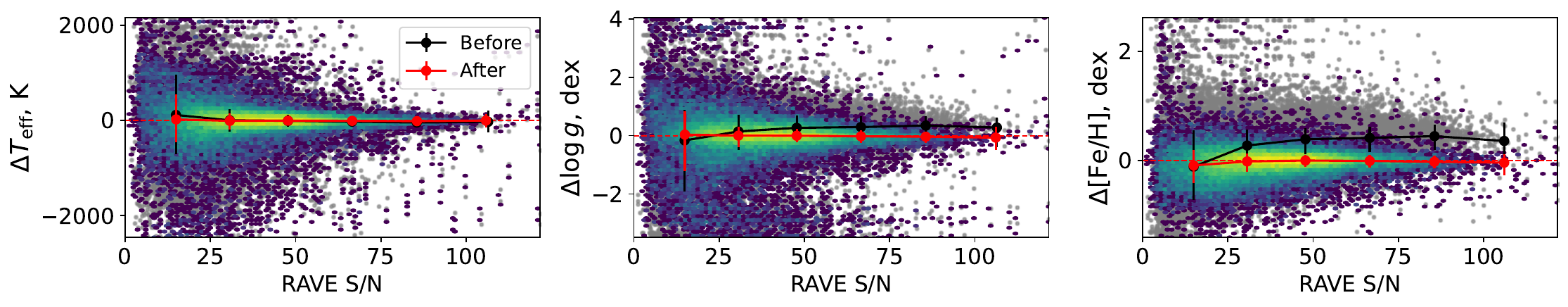} \\
        \includegraphics[width=0.8\linewidth]{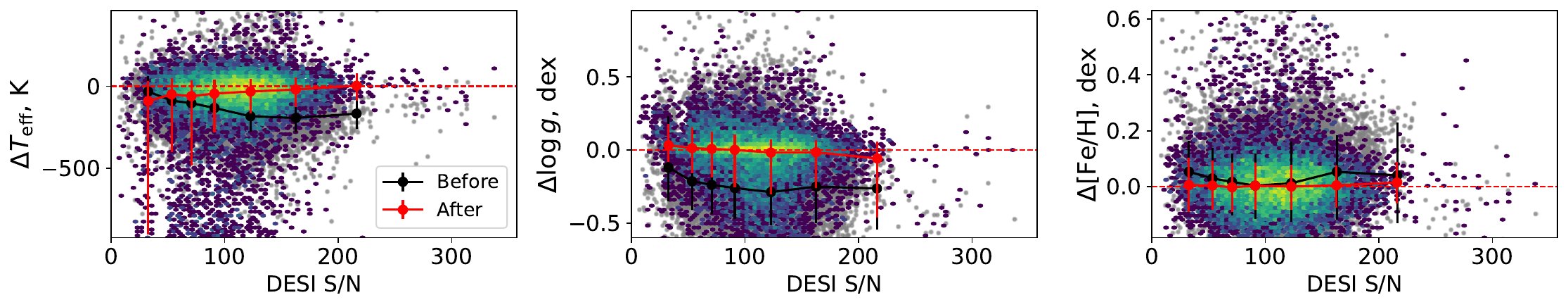} \\
        \includegraphics[width=0.8\linewidth]{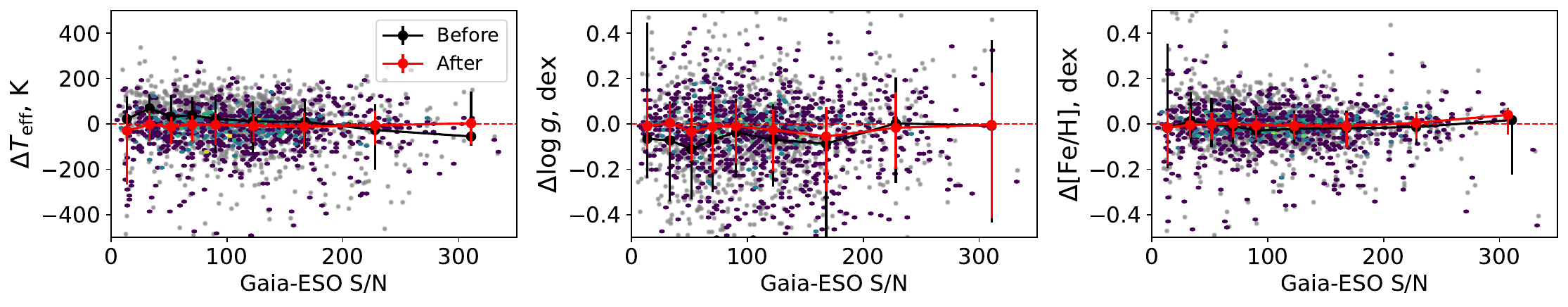} \\
        \includegraphics[width=0.8\linewidth]{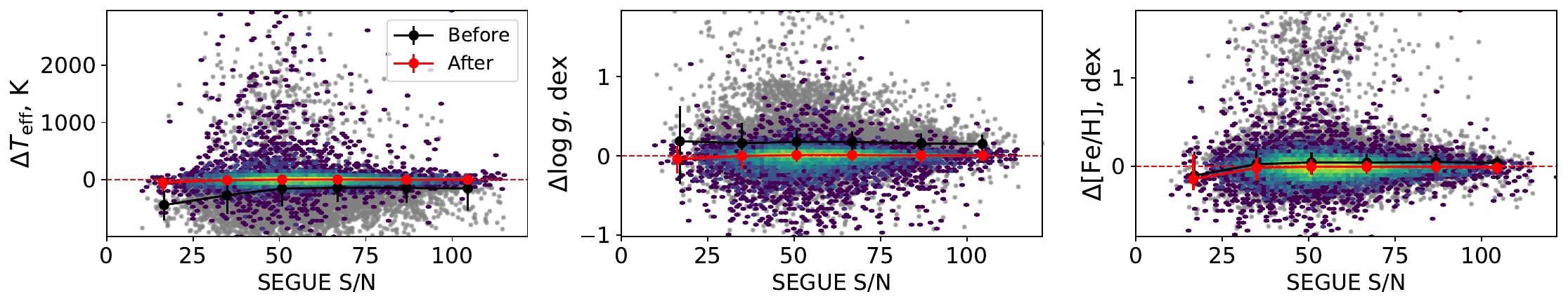}
        
        \vspace{-0.5em}
    \caption{Differences between the stellar parameters from each survey and the corresponding main reference scale, APOGEE or GALAH, as a function of S/N. The original differences are shown as grey points with their median trend in black, while the corrected differences are shown in colour with the median trend in red.}
    \label{fig:snr_all}
\end{figure*}

Since S/N varies strongly within and between surveys, we also check whether the calibrated residuals depend on S/N (Fig.~\ref{fig:snr_all}). After calibration, the S/N-dependent trends are generally strongly reduced, with the median residual becoming approximately flat and centred around zero in most cases. The main exception is [Fe/H] for Gaia RVS, where the improvement is more modest and a residual dependence on S/N remains. For $\log g$ in Gaia RVS, the trend also persists after calibration, but its amplitude is substantially reduced.

\section{External calibration on PASTEL}
\label{app:pastel_calibration}

We constructed a consolidated external reference sample from the PASTEL \citep{Soubiran2010, Soubiran2016}
catalogue\footnote{\url{https://vizier.cds.unistra.fr/viz-bin/VizieR?-source=B/pastel}}.
A minimal cleaning was applied before combining the literature measurements.
Measurements from some sources were excluded because, for the same stars, they were systematically inconsistent with the rest of the compilation. Two entries from \citep{Soto2018cat} with $T_{\rm eff}=3500$~K were removed, since these cases are flagged in the original catalogue as entries for which convergence was not reached.

Then, the remaining PASTEL measurements were grouped by object identifier and processed
independently for each $T_{\rm eff}$, $\log g$, and
[Fe/H]. Objects with a single measurement were retained only when a quoted
uncertainty was available. For objects with multiple measurements, exact
duplicates, defined as entries with identical values and uncertainties, were
removed before computing the adopted reference value.

Missing uncertainties were imputed only for objects with multiple
measurements, so that the scatter between independent literature values could
still be used as an additional consistency check. The adopted imputation
rules were based on the comparison between PASTEL and the \textit{Gaia} benchmark
stars. For $T_{\rm eff}$, missing uncertainties were set to 120~K for
$T_{\rm eff}<6000$~K and to $T_{\rm eff}/10$ otherwise. For $\log g$, they
were set to 0.4~dex for $\log g<3.5$ and to 0.1~dex otherwise. For [Fe/H],
they were set to 0.15~dex for [Fe/H]$<-0.5$ and to 0.07~dex otherwise.

We also imposed a simple internal-consistency requirement. For each object
and parameter, the group of measurements was discarded if the standard
deviation exceeded 20\% of the mean value in $T_{\rm eff}$, or 0.5~dex in
either $\log g$ or [Fe/H].

For objects with two or more usable measurements, the adopted reference value
was computed as an inverse-variance weighted mean,
\[
w_i = \frac{1}{\sigma_i^{2}},
\qquad
\bar{x} = \frac{\sum_i w_i x_i}{\sum_i w_i}.
\]
The uncertainty on the weighted mean was estimated by combining the formal
inverse-variance term with the observed scatter between measurements,
\[
\sigma_{\bar{x}} =
\sqrt{
\frac{1}{\sum_i w_i}
+
\frac{\sum_i w_i (x_i-\bar{x})^2}{(\sum_i w_i)(n-1)}
}.
\]

After this filtering, the final PASTEL reference sample contains 25\,976
unique stars. The effective temperatures range from about 4000 to
22\,000~K, with a median value of 6296~K. Surface gravities span
0.6--4.7~dex, and metallicities range from $-3.2$ to $0.4$~dex. The median
uncertainties are about 76~K in $T_{\rm eff}$, 0.1~dex in $\log g$, and
0.06~dex in [Fe/H].

\begin{figure*}[t]
    \centering
    \includegraphics[width=0.8\textwidth]{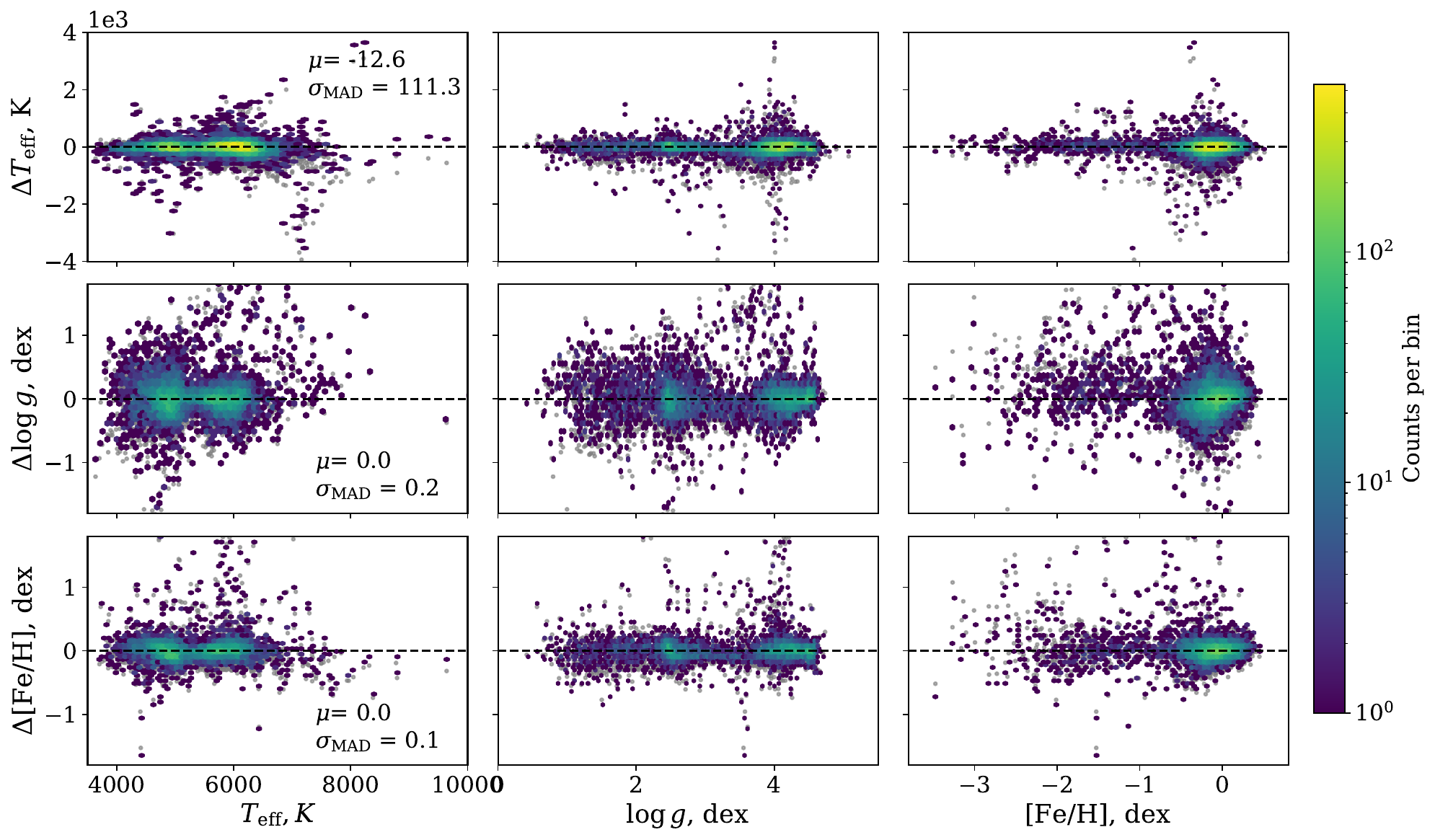}
    \caption{Residual differences with respect to PASTEL after the external
    calibration. The residuals in $T_{\mathrm{eff}}$, $\log g$, and [Fe/H]
    are shown as a function of the atmospheric parameters to assess the
    remaining systematic trends after applying the linear corrections. The grey points show the residuals before the external calibration.}
    \label{fig:pastel_calib_trends}
\end{figure*}

For each atmospheric parameter, we modelled the residual between SoS catalogue values and
PASTEL as a linear function of the stellar parameters. This final step
corrects remaining global offsets and simple parameter-dependent trends
while keeping the external calibration transparent. The resulting correction
relations are

\begin{align}
T_{\rm eff,corr} &= -732.5 + 1.1683\,T_{\rm eff} - 23.38\log g \nonumber\\
  &\quad + 15.65\,[{\rm Fe/H}], \label{eq:teffcorr}\\
\log g_{\rm corr} &= 0.2557 + 1.526\times10^{-5}\,T_{\rm eff} + 0.9261\log g \nonumber\\
  &\quad + 0.1584\,[{\rm Fe/H}], \label{eq:loggcorr}\\
[{\rm Fe/H}]_{\rm corr} &= -0.006085 + 3.574\times10^{-5}\,T_{\rm eff} - 0.03785\log g \nonumber\\
  &\quad + 1.0888\,[{\rm Fe/H}]. \label{eq:fehcorr}
\end{align}

These relations are applied to the internally homogenised parameters to
obtain the final externally calibrated catalogue values.
Figure~\ref{fig:pastel_calib_trends} shows the residual trends after this
external calibration as a function of the stellar parameters.

The uncertainties of the internally homogenised parameters are propagated
through the external calibration using the same Monte Carlo procedure as in
the previous calibration steps. The final uncertainty intervals are defined
from the sampled distributions and are reported in the catalogue as lower
and upper bounds.

The external calibration corrects parameter-dependent
systematic offsets but does not reduce the star-to-star dispersion, which reflects the random uncertainties of both catalogues. For the 14\,360
stars in common with the PASTEL reference sample, this dispersion
is consistent with the combined uncertainties of the two catalogues:
the normalised residuals,
$\Delta/(\sigma_{\rm SoS}^2 + \sigma_{\rm PASTEL}^2)^{1/2}$,
have $\sigma_{\rm MAD} = 0.97$, 1.15, and 0.92 for $T_{\rm eff}$,
$\log g$, and [Fe/H], respectively. The excess in $\log g$
is concentrated in giants ($\sigma_{\rm MAD} = 1.24$ for
$\log g < 3.5$), for which the quoted PASTEL uncertainties
(median $\sim$0.2\,dex) are smaller than the $\sim$0.4\,dex
scatter of PASTEL values in comparison to the \textit{Gaia} benchmark
stars.


\end{document}